\documentclass[12pt]{article}
\usepackage[margin=1in]{geometry}
\usepackage{amsmath, amssymb}
\usepackage{natbib}
\usepackage[colorlinks=true, linkcolor=black, citecolor=blue, urlcolor=blue]{hyperref}
\usepackage{setspace}
\usepackage[dvipsnames]{xcolor}
\usepackage{enumitem}
\usepackage[most]{tcolorbox}

\definecolor{ink}{HTML}{1A1A1A}
\definecolor{accent}{HTML}{1F4E79}
\definecolor{soft}{HTML}{5B6B7B}
\definecolor{rule}{HTML}{C9D3DD}
\definecolor{forkbg}{HTML}{EAF0F6}

\newcommand{\cksection}[1]{%
  \par\vspace{11pt}%
  {\normalfont\bfseries\color{accent}\large #1}\par\vspace{2pt}%
  {\color{rule}\hrule height 0.8pt}\par\vspace{4pt}}

\newcommand{\cksubsection}[1]{%
  \par\vspace{8pt}{\normalfont\bfseries\color{accent} #1}\par\vspace{2pt}}

\providecommand{\cknote}[1]{\par\vspace{1pt}{\color{soft}\itshape\small #1}\par\vspace{4pt}}

\newlist{checkitems}{itemize}{1}
\setlist[checkitems]{leftmargin=1.6em,itemsep=2.5pt,topsep=2pt,parsep=0pt,
  label=\textcolor{accent}{$\square$}}

\newlist{subitems}{itemize}{1}
\setlist[subitems]{leftmargin=1.5em,itemsep=1.5pt,topsep=2pt,parsep=0pt,
  label=\textcolor{soft}{\textendash}}

\usepackage{tikz}
\usetikzlibrary{arrows.meta}

\title{The Measurement Revolution? Credible Measurement and Inference in the Age of AI\thanks{
This review accompanies the National Bureau of Economic Research 2026 Methods Lecture on \textit{Estimation and Inference with AI-Generated Data}, available at \url{https://www.nber.org/conferences/si-2026-methods-lecture-estimation-and-inference-ai-generated-data}.}}
\author{Melissa Dell and Ashesh Rambachan}
\date{August 2026}

\begin{document}

\maketitle

\begin{abstract}
Artificial intelligence (AI) is transforming measurement in economics. AI models convert unstructured data, such as text and images, into structured variables at low cost, making previously prohibitive measurement feasible at scale. This shifts the bottleneck from finding \textit{any} scalable measure of a phenomenon to choosing among \textit{many} plausible ones, which may support different empirical conclusions. This review provides guidance for navigating that shift. We describe three stages at which AI enters the measurement pipeline---discovery, construct definition, and observation---and what each demands of researchers. We argue that credible inference with AI-generated variables requires appropriately designed validation: anchoring measurement to explicit criteria, rather than informal claims that a proxy is reasonable. We then examine how validation samples support valid inference even when AI predictions are arbitrarily biased, and what can be done when a random validation sample is unavailable.
\end{abstract}

\clearpage

\section{Introduction}

Artificial intelligence (AI) is transforming measurement in economics. 
AI models can convert unstructured data (\textit{e.g.,} text, images, audio, and video) into low-dimensional structured variables for economic analysis at very low marginal cost. 
For example, large language models (LLMs) can code open-ended survey responses, historical documents, and massive-scale web text corpora, while computer vision models can transform satellite imagery into measures of poverty, deforestation, and pollution at fine spatial resolution and with broad geographic coverage. 
AI makes measurement endeavors that were once prohibitively costly feasible at scale, allowing economists to revisit long-standing questions with new evidence and to study questions that were previously out of reach.

Although measurement is often conceived as passive observation through an instrument---whether a ruler or an LLM---in economics it is better understood as a disciplined simplification of a complex, high-dimensional world. Jorge Luis \citet{Borges1998Exactitude} vividly captures this in ``On Exactitude in Science,'' where he imagines cartographers who pursue perfect representation until they produce a map ``whose size was that of the Empire.'' The perfect map, however, is useless: because it no longer reduces complexity, it ceases to function as a map. Our goal is not to produce a perfect measure of economic reality, but to construct a simplified representation that reveals the broader landscape and helps navigate it.

More formally, measurement specifies and implements a function $m(\cdot)$ that projects high-dimensional reality $U$ into a lower-dimensional representation $m(U)$ that can be interpreted and analyzed. For example, economists have mapped newspaper articles (high-dimensional texts) into an indicator for whether the article discusses economic policy uncertainty \citep{baker2016measuring}.
Because $U$ is high-dimensional, there are often many possible ways to measure the same concept. Different measurement functions $m_j(\cdot)$ may preserve different features of $U$ and may therefore support different empirical conclusions.

Deep neural networks are the foundation of modern AI, and one of their most common uses is as flexible implementations of measurement functions \citep[for a survey aimed at economists, see][]{Dell2025}. 
Earlier approaches to unstructured data often encoded human judgment directly through hand-engineered rules. 
Neural networks instead learn from empirical examples, adjusting millions or billions of parameters to perform a specified task, such as predicting the next word or classifying an image. 
Given sufficient training data and an appropriate training objective, they learn parameters that compress $U$ into a lower-dimensional representation to perform the task at hand, without requiring the researcher to specify which features should be preserved, discarded, or transformed. The representations they learn can often be reused for other tasks, a principle known as ``transfer learning.''

Because neural networks are highly scalable, they can dramatically lower the marginal cost of implementing measurement functions $m(\cdot)$, extending systematic measurement to domains that were previously out of reach. 
The size of the overall cost reduction depends on the task. In some settings, AI measurement requires substantial fixed investments: to achieve sufficiently accurate predictions, researchers need to develop high-quality training data and fine-tune customized models, so constructing $m(\cdot)$ remains costly although applying it at scale is cheap. In other settings, off-the-shelf or lightly adapted models, such as commercial LLMs, perform sufficiently well to dramatically lower both fixed and marginal costs.

Cheap measurement makes it feasible to construct many alternative measures of the same underlying concept, and AI introduces many implementation choices that can meaningfully affect the resulting measure. Researchers can vary the measurement rubric, model, and prompt; when fine-tuning, they must also choose the training-data distribution and tuning strategy. Model parameters add further degrees of freedom, even when defaults are available. 

This abundance of potential measures marks a substantial shift in empirical work. 
Researchers have often had to rely on externally constructed measures as fixed proxies for theoretically meaningful concepts---for example, using an external measure of expropriation risk for foreigners as a proxy for property rights institutions. 
Such proxies may contain arbitrary measurement error or poorly align with the construct of interest, but they were often the only scalable option.
With AI, the bottleneck is shifting from finding \textit{any} scalable measure of a phenomenon to choosing among \textit{many} plausible ones. 

Interpreting differences across abundant AI measures is complicated. 
While extremely powerful, deep neural networks are in many ways black boxes. 
When measurement is delegated to an AI model, the features that are preserved in $m(U)$ are determined implicitly through complex interactions among millions or billions of learned parameters. 
Direct inspection of these parameters generally does not reveal an interpretable understanding of how the model constructs $m(U)$. 
This opacity may not matter when the end goal is prediction \citep[][]{MullainathanSpiess2017}, but it becomes consequential when AI-generated variables are used in downstream empirical analysis \citep[][]{LudwigMullainathanRambachan2026}. 
Economists seek to interpret relationships involving $m(U)$, which requires understanding what measures capture and why alternative measures of the same concept can yield different conclusions, a phenomenon documented by a growing literature reviewed in this article.

These considerations translate into several concrete challenges for incorporating AI-generated measures into empirical analysis.
The researcher must choose which measure(s) to use and report, and post-selection inference concerns can arise \citep[e.g.,][]{baumann2025large}. 
This is compounded by the fact that researchers can generate large volumes of noisy, poorly validated, or difficult-to-interpret measures, what is colloquially called ``AI slop.'' 
Researchers must decide when costly investments in training data, fine-tuning, or model refinement are worthwhile to improve poorly performing measures. 
This is difficult to assess without a principled framework for how measurement quality affects estimates of the target parameter. 
Finally, reproducibility concerns arise because proprietary models are frequently deprecated \citep[e.g.,][]{BarrieEtAl24-LLMReplication, coqueret2026randomnesslargelanguagemodels}.

How should economists approach AI measurement given these challenges? 
To elucidate this question, this review first describes the three stages at which AI can enter the measurement pipeline, and what each stage demands of the researcher (Section~\ref{section:pipeline}). 
\textit{Discovery} selects which features of the world to measure at all. 
\textit{Construct definition} fixes what the selected concept means, in terms precise enough to apply to a specific setting. 
\textit{Observation} implements that definition at scale, producing the variable that enters the analysis. 
AI can propose candidate features directly from the data, turn a single concept into many competing operationalizations, and apply a definition to millions of units at minimal marginal cost. 

The existence of many potential measures does not make measurement inherently subjective.
Rather, it underscores the importance of validating whether measurements satisfy clearly articulated objectives, which are themselves subject to scientific debate. 
Credible inference with AI-generated predictions requires appropriately designed measurement validation.
We define validation as anchoring measurement to explicit, observable criteria, rather than simply relying on informal arguments that a proxy is reasonable, and discuss the evidence for its importance (Section~\ref{sec:credible}). 
Diverse literatures from econometrics, statistics, biostatistics, machine learning, psychometrics, and computational social science provide a well-developed foundation for designing validation, while also highlighting important gaps in existing methods.

Statistical methods are the most developed for the observation stage of measurement, and this is where AI is making the most rapid inroads. 
Hence, we largely focus on observation in this review. 
We examine how validation samples can support valid inference even when AI predictions are arbitrarily biased (Section~\ref{obs:mar}). 
The key is that prediction errors are learned from the validation design, rather than inferred from assumptions about how the model generates its outputs. 
A random validation sample, however, is not always available. 
Credibility then rests on assumptions about how AI models behave across settings, rather than on the design of the validation sample (Section~\ref{sec:noval}). Those assumptions can be stated precisely, and some can be checked against evidence on how frontier models err. However, this remains a largely open frontier given the enormous complexity of deep neural networks. 

The opportunities created by AI, and the credibility challenges that accompany them, echo an earlier transformation in economics. 
When personal computing made estimation cheap in the 1990s, researchers could run not just one empirical specification but hundreds, thousands, or even millions \citep{sala1997just}. 
That forced the field to ask what should discipline the choice among many possible specifications.
The response was the credibility revolution \citep[][]{Angrist2010CredibilityRevolution, Angrist2022}: rich literatures developed around causal inference, robustness, and multiple hypothesis testing, while empirical practice increasingly emphasized specifications designed to support clear interpretation.

AI moves specification search upstream, from choosing among empirical specifications to choosing among alternative measurements of a high-dimensional reality.
The relevant lesson is analogous: expanding measurement choice highlights credibility problems that require empirical discipline and transparency to address.
Researchers should specify verifiable measurement aims, evaluate how well their measurements achieve those aims, and design measurement to support valid, interpretable inference.
This article synthesizes both classic and emerging literatures that build the foundation for credible measurement in the AI era.

\section{The Measurement Pipeline: Discovery, Definition, and Observation}\label{section:pipeline}

Colloquially, measurement is often viewed as the act of observing a quantity with an instrument or procedure. For economic research, it is more useful to view measurement as a pipeline with three stages. First, \emph{discovery}: which features of the world should the researcher measure? Second, \emph{construct definition}: how should the relevant concept be operationally defined, given the research question? Third, \emph{observation}: how can that definition be implemented at scale? A researcher might, for example, identify economic policy uncertainty as the feature of interest, write a detailed rubric specifying what counts as economic policy uncertainty and what does not, and implement that rubric by prompting an LLM or designing a keyword query over economic news articles \citep{baker2016measuring}.\footnote{Statistical inference is closely related: it maps structured data into a lower-dimensional representation, such as a target parameter, and could itself be viewed as a stage of the measurement process. We reserve measurement for the construction of variables from underlying reality, and inference for the use of those variables to learn about a target parameter.}

Each stage carries a different kind of uncertainty, connects to different methodological literatures, and calls for a different form of validation. 
Discovery involves uncertainty about the structure of the world: which patterns, dimensions, or relationships exist and are worth measuring? 
It has historically been guided by theory and exploratory analysis, and it requires evidence that a procedure has uncovered meaningful structure. 
Construct definition involves conceptual uncertainty: how should a particular phenomenon be translated into an operational construct? 
It draws again on theory and domain expertise but has also spurred construct validity literatures in psychometrics and survey design. It requires evidence that an operationalization captures the intended concept. 
Finally, observation involves statistical uncertainty from imperfect implementation: how accurately can a construct be measured at scale? It draws on the measurement error and semiparametric inference literatures, and it requires evidence that a procedure implements the chosen criterion.

AI is transforming all three stages of the measurement pipeline.
In discovery, it can surface patterns that no one thought to specify in advance.
In construct definition, it can lower the cost of generating and comparing alternative operationalizations of a concept. 
Finally, in observation, it can apply a chosen definition at scale. 
Figure~\ref{fig:pipeline} summarizes the three stages, what each requires, and how AI is entering each of them. 
This section introduces each stage in turn. 
The remainder of the article then concentrates on observation, where the methods for working with AI-generated measurements are the most developed, and AI use is accelerating most rapidly.

\definecolor{pipeink}{HTML}{1F2933}
\definecolor{pipemuted}{HTML}{7B8794}
\definecolor{piperule}{HTML}{DCE2E8}
\definecolor{pipebg}{HTML}{F7F9FB}
\definecolor{stageA}{HTML}{3D6A9E}   
\definecolor{stageB}{HTML}{2F8A7A}   
\definecolor{stageC}{HTML}{A8642A}   

\newcommand{\pipeperson}[2]{
  \begin{scope}[shift={#1}]
    \fill[#2] (0,0.19) circle (0.105);
    \fill[#2, rounded corners=1pt]
      (-0.175,-0.22) -- (-0.175,-0.03) arc (180:0:0.175) -- (0.175,-0.22) -- cycle;
  \end{scope}}
\newcommand{\piperobot}[2]{
  \begin{scope}[shift={#1}]
    \draw[#2, line width=0.8pt] (0,0.18) -- (0,0.30);
    \fill[#2] (0,0.325) circle (0.04);
    \fill[#2, rounded corners=2.2pt] (-0.20,-0.19) rectangle (0.20,0.19);
    \fill[white] (-0.085,0.04) circle (0.038);
    \fill[white] ( 0.085,0.04) circle (0.038);
    \fill[white, rounded corners=0.4pt] (-0.075,-0.11) rectangle (0.075,-0.08);
  \end{scope}}

\newcommand{\pipecard}[8]{%
  \fill[pipebg, rounded corners=3pt] (#1,0) rectangle (#1+5,-6.75);
  \begin{scope}
    \clip[rounded corners=3pt] (#1,0) rectangle (#1+5,-6.75);
    \fill[#2] (#1,0) rectangle (#1+5,-0.86);
    \fill[#2!8] (#1,-5.20) rectangle (#1+5,-6.75);
  \end{scope}
  \draw[piperule, line width=0.5pt, rounded corners=3pt] (#1,0) rectangle (#1+5,-6.75);
  \fill[white] (#1+0.44,-0.43) circle (0.185);
  \node[font=\sffamily\tiny\bfseries, text=#2] at (#1+0.44,-0.43) {#3};
  \node[anchor=west, font=\sffamily\scriptsize\bfseries, text=white]
       at (#1+0.76,-0.43) {#4};
  \node[anchor=north west, text width=4.44cm, align=left,
        font=\sffamily\scriptsize, text=pipeink, inner sep=0pt]
       at (#1+0.28,-1.12) {#5};
  \draw[piperule, line width=0.5pt] (#1+0.28,-2.28) -- (#1+4.72,-2.28);
  \pipeperson{(#1+0.55,-3.02)}{pipemuted}
  \node[anchor=north west, font=\sffamily\scriptsize\bfseries, text=pipemuted,
        inner sep=0pt] at (#1+1.02,-2.41) {Before AI};
  \node[anchor=north west, text width=3.70cm, align=left,
        font=\sffamily\scriptsize, text=pipemuted, inner sep=0pt]
       at (#1+1.02,-2.75) {#6};
  \draw[piperule, line width=0.5pt] (#1+0.28,-3.74) -- (#1+4.72,-3.74);
  \piperobot{(#1+0.55,-4.48)}{#2}
  \node[anchor=north west, font=\sffamily\scriptsize\bfseries, text=#2,
        inner sep=0pt] at (#1+1.02,-3.87) {With AI};
  \node[anchor=north west, text width=3.70cm, align=left,
        font=\sffamily\scriptsize, text=pipeink, inner sep=0pt]
       at (#1+1.02,-4.21) {#7};
  \draw[piperule, line width=0.5pt] (#1+0.28,-5.20) -- (#1+4.72,-5.20);
  \node[anchor=north west, font=\sffamily\tiny\bfseries, text=#2, inner sep=0pt]
       at (#1+0.28,-5.39) {VALIDATION MUST SHOW};
  \node[anchor=north west, text width=4.44cm, align=left,
        font=\sffamily\scriptsize, text=pipeink, inner sep=0pt]
       at (#1+0.28,-5.71) {#8};}

\begin{figure}[t]
\centering
{\sffamily\hyphenpenalty=10000\exhyphenpenalty=10000\relax
\begin{tikzpicture}

\draw[pipemuted, line width=1.0pt, -{Latex[length=2.6mm,width=2.0mm]}]
      (0.15,0.60) -- (15.85,0.60);
\node[anchor=west, font=\sffamily\scriptsize\itshape, text=pipemuted]
     at (0.15,0.92) {High-dimensional reality $U$};
\node[anchor=east, font=\sffamily\scriptsize\itshape, text=pipemuted]
     at (15.85,0.92) {Measured variable $m(U)$ for analysis};

\pipecard{0}{stageA}{1}{DISCOVERY}
  {\textbf{Goal.} Determine which features of the world are worth measuring.}
  {Theory and exploratory analysis propose candidates.}
  {Procedures surface patterns no one thought to specify.}
  {That the procedure uncovered meaningful structure on out-of-sample data.}

\pipecard{5.5}{stageB}{2}{CONSTRUCT DEFINITION}
  {\textbf{Goal.} Translate that feature into an operational construct.}
  {Theory, domain expertise, and construct validity.}
  {Evaluating rival definitions is much cheaper.}
  {That the construct definition captures the intended concept.}

\pipecard{11}{stageC}{3}{OBSERVATION}
  {\textbf{Goal.} Implement the construct definition at scale.}
  {Trained annotators or field teams, at the scale the budget allows.}
  {A model applies the definition to potentially large-scale data.}
  {How well the procedure implements the chosen criterion.}

\foreach \x in {5.25,10.75}{
  \fill[pipemuted] (\x-0.10,-3.07) -- (\x+0.13,-3.37) -- (\x-0.10,-3.67) -- cycle;}

\end{tikzpicture}}
\caption{\textbf{The measurement pipeline.} Measurement proceeds in three
stages: discovery, construct definition, and observation. AI is transforming each stage, creating both opportunities and challenges for empirical research.}
\label{fig:pipeline}
\end{figure}
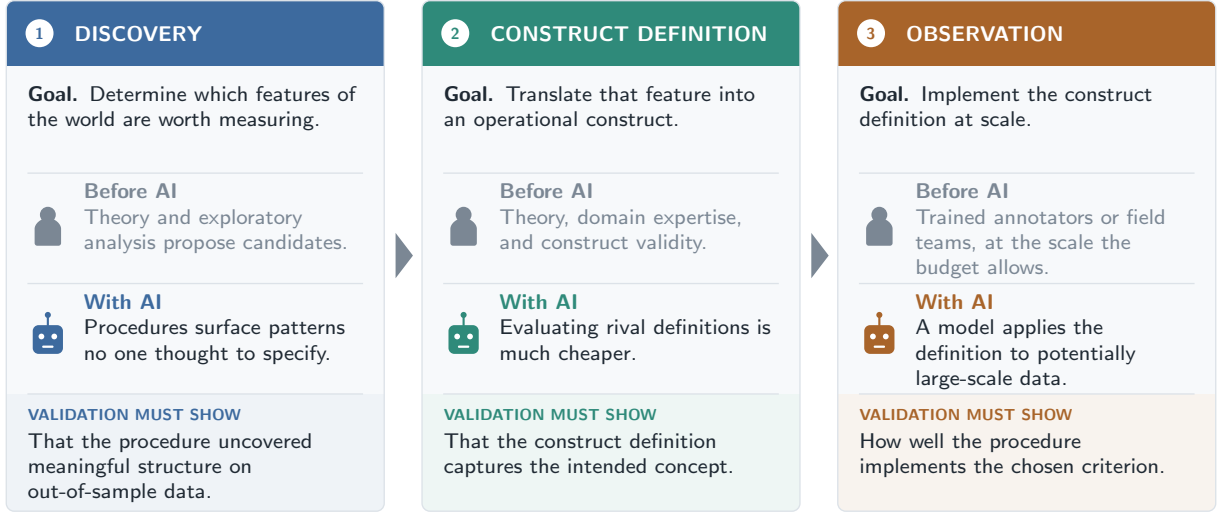

\subsection{Discovery}\label{subsec:disc}

\textit{Discovery} has typically been guided by theory, contextual knowledge, or incremental advances in an existing literature. This is productive when a problem is well enough understood that pre-specified measurement is warranted. But it also means that the structured datasets anchoring empirical economics contain only the features that someone previously decided to measure: the closed survey items we field, the outcomes we track in a randomized experiment, the variables of an administrative extract. Unstructured data record much richer information. Relying on researcher intuition and creativity to decide what to measure in these data can produce a bottleneck, slowing what we can discover.

In data-driven discovery, the researcher specifies a procedure rather than a predefined construct. When the researcher has a particular outcome in mind, these procedures can be directed toward uncovering constructs that relate to it. This activity is commonly known as ``hypothesis generation.'' 
The resulting structure enters measurement as an input to construct definition, where it is refined into a concept and then evaluated in a separate sample, or it may itself serve as the target measurement.

These procedures are not hypothetical: across modalities, they have already produced validated discoveries. \citet{LudwigMullainathan2024} turn the camera on the judge, predicting pretrial detention decisions from defendants' mug shots. Much of the mug shot's predictive signal is unexplained by structured characteristics or known facial features: something unnamed in the image systematically drives detention decisions. Their hypothesis generation procedure names two novel hypotheses, well-groomed and heavy-faced, that survive validation on held-out data. \citet{BaronEtAl2026} bring the same logic to text, using the rich case notes written by child protective services investigators to generate hypotheses about what separates high-performing investigators from the rest. \citet{ObermeyerEtAl2026} extend it to medical waveforms, morphing electrocardiograms along a clinical predictor to surface a novel feature of the waveform that predicts sudden cardiac death.

The broader literature on hypothesis generation is rapidly growing, spanning economics, computer science, and beyond \citep[see][for a brief overview]{MullainathanRambachanScience2025}. Applications now include effective teaching \citep{Workman2025}, medical notes \citep{DonahueEtAl2026}, and cognitive experiments \citep{ZhuEtAl2026}.
Hypothesis generation techniques also enable researchers to collect and then analyze open-ended survey responses: AI-conducted interviews make open-ended elicitation cheap at scale \citep{HaalandEtAl2024, ChopraHaaland2026, GeieckeJaravel2026}, and hypothesis generation procedures can automate the discovery of recurring themes in the resulting responses \citep{WangEtAl2026}.
An accompanying methodological literature on hypothesis generation studies how to describe the differences between text corpora \citep{ZhongEtAl2022, ZhongEtAl2023}, generate hypotheses with large language models \citep{ZhouEtAl2024, BatistaRoss2024} and sparse autoencoders \citep{MovvaEtAl2025, MovvaEtAl2026, PengEtAl2025}, conduct statistical inference on generated hypotheses \citep{ModarressiEtAl2025, Carlson2026}, and benchmark hypothesis generation procedures \citep{LiuEtAl2025}.

Discovery is disciplined not by how hypotheses are generated but by how they are evaluated. Evidence on the quality of a generated hypothesis must come from held-out data: the researcher measures the named construct there and asks how well it predicts the outcome of interest. If the generating procedure never touches that held-out sample, it does not matter how the hypothesis was produced, and procedures as different as an expert's intuition and a large language model can be judged by the same standard. This is the logic of the ``common task framework,'' which has organized much of the progress in modern machine learning \citep{Donoho2024}.

Nonetheless, a generated hypothesis is only the beginning. Discovery yields a named concept. To study that concept, for example estimate its prevalence or include it in a regression, the researcher must define it precisely and measure it across the full dataset. Those are the tasks of the next two stages.

\subsection{Construct Definition}

\textit{Construct definition} asks how a given concept should be operationalized in a particular research setting. 
This stage is difficult because many concepts of interest are multidimensional, not directly observable, and closely relate to other phenomena. 
A measure of trust, for example, should capture the dimension of trust relevant to the research question without simply reproducing institutional quality, income, or other correlated variables. Theory and domain expertise therefore remain central: they supply the criteria against which an operationalization can be judged.

AI makes it cheap to generate operationalizations themselves. A researcher can ask a model to draft many rubrics for a concept like economic policy uncertainty, each plausible, each drawing the concept's boundary somewhere slightly different. The researcher is then choosing among constructs, not merely among ways of implementing one.

Adjudicating among such candidates is an old problem, and the classic treatment is \citet{cronbach1955construct}. Many social science concepts, such as trust, social capital, and intelligence, are latent and complex. Since they cannot be directly observed, \citeauthor{cronbach1955construct} argue that validation must proceed \textit{indirectly} through a ``nomological network'': theory or contextual knowledge that implies what the construct should correlate with (referred to as convergent validity), what it should be distinct from (referred to as discriminant validity), and how stable it should be across settings. 
A construct is validated not by any single observable variable but by a broader pattern of relationships.
In the language of econometrics, this logic is analogous to partial identification under a set of theoretically motivated moment restrictions: no single restriction establishes that an operationalization captures the intended construct, but additional restrictions narrow the set of plausible measures.

Construct validity is increasingly important for economics. 
The discipline has historically emphasized measures that are directly observed or for which direct validation is plausible, such as prices, schooling, and employment, devoting less attention to construct validity than work in psychometrics or survey design. Where such constructs are well established, construct validity is largely settled. 
But economists increasingly study complex latent concepts such as economic policy uncertainty, trust, and institutional quality, where construct validity is far more contested.

The abundance of candidate operationalizations is also an opportunity. 
When many are available, the researcher must say what distinguishes them, which forces the construct and its nomological network to be stated precisely enough to adjudicate among candidates. 
Where off-the-shelf models perform well enough, AI lowers the cost of that adjudication too, since the observable implications of the network, such as related outcomes, correlates, contrasts, and contextual patterns, can themselves be measured cheaply. 
What binds is not the supply of candidate measures but the theory and domain knowledge needed to say precisely what we are trying to measure. 
AI can draft rubrics and measure their implications; it cannot tell the researcher which operationalization answers the research question. 
That judgment rests on theory-driven measurement and substantive expertise about the setting. 
If the theoretical framework is weak, a measure's failure to align with the network may indict the underlying theory rather than the measure. 

When credible external criteria are unavailable altogether, it may be better to focus on the discovery stage. 
Data-driven discovery can surface interpretable patterns without requiring strong priors about constructs that are not yet well understood. 
Those discovered patterns can then inform theory and construct definition.

\subsection{Observation}

In \textit{observation}, AI serves as a scalable measurement technology. 
Once a researcher has specified an operational definition, an AI system can apply it repeatedly at low marginal cost, for example classifying large text corpora or extracting information from vast collections of images at a scale that would be infeasible for human coders. 

Where construct validity asks whether an operationalization captures the concept needed for the research question, observational validity asks whether a procedure implements that operationalization. 
To understand this difference, a useful, though imperfect, analogy is the distinction between internal and external validity: observational validity (like internal validity) assesses credibility conditional on a specified target, while construct validity (like external validity) asks whether that target answers the broader question motivating the research. 
The criterion may itself be a flawed operationalization of the concept of interest, but it is explicit and interpretable. Observation takes that criterion as fixed and asks only whether a procedure implements it faithfully.

Errors from AI-generated measurements are unlikely to be classical. 
Systematic biases can arise from network architecture, the distribution of training data, and other implementation details, and the nonlinear transformations applied at each layer of the neural network, together with the frequent use of binary or multiclass outputs, violate classical measurement error assumptions. 
These biases do not stay in the measurement; they propagate to estimates of the target parameter. 
AI-generated variables therefore cannot simply be substituted for the constructs they are meant to capture. 

The remainder of the article takes up what to do instead: Section~\ref{sec:credible} asks what credible measurement with AI-generated variables requires, and Sections~\ref{obs:mar} and~\ref{sec:noval} give the two answers, depending on whether the researcher can construct a random validation sample.

\section{From AI Outputs to Credible Measurements}\label{sec:credible}

\subsection{Validation as an Anchor for Credible Measurement}

Intuitively, validation plays an important role in establishing credibility, but what should be validated? 
Validation cannot show that a constructed measure captures a latent concept perfectly or recovers ``fundamental truth.'' Measurement aims at simplified representations of a complex reality that preserve the features needed to draw useful inferences. Ideally, validation would therefore establish that a measure achieves that broader aim. 
This is often extremely challenging to assess quantitatively, because the underlying reality is complex, partially unobservable, multi-causal, and not easy to manipulate experimentally. 

When a measure is treated as a proxy, it is common to argue (often qualitatively) why it achieves this aim. But qualitative arguments often cannot adjudicate among many plausible measures that could lead to different conclusions. Consider economic policy uncertainty. One researcher counts newspaper articles containing terms for the economy, policy, and uncertainty; another asks a language model whether an article conveys uncertainty about economic policy; a third restricts attention to federal policy alone. 
Each choice can be defended as a reasonable proxy, and each may produce a different series. 

Validation therefore has a simpler---and more achievable---empirical aim. 
It requires specifying observable measurement criteria, assessing how candidate measures behave relative to those criteria, and adjusting target estimates for systematic errors in measurement. 
In the economic policy uncertainty example, it means committing to a rubric---supported by theory and domain expertise---that states what counts as expressing economic policy uncertainty and what edge cases do not. 
Competing measures can then be scored against that rubric, disagreements traced to particular articles, and systematic error corrected. Such criteria are themselves imperfect, but they can be stated explicitly and refined. 
This encourages scientific debate and incremental improvements in measurement for difficult-to-measure concepts.

Anchoring measurement to \textit{explicit criteria} is especially important in the age of AI, because these criteria address a central limitation of modern AI systems. 
These systems are built from neural networks whose mappings from inputs to outputs are too complex to understand by inspecting their parameters, and whose learned criteria are not expressed as explicit, researcher-auditable rules. 
Their biases can shift across inputs in ways that humans predict poorly \citep{VafaEtAl2024}. 
Validation makes the black box of modern AI interpretable, not by elucidating its internal workings, but by evaluating its outputs against criteria specified by the researcher.

Validation also addresses the abundance of measures that AI can cheaply produce. 
When many candidate measures can be constructed, measurement choices themselves become objects of empirical comparison. 
Validation disciplines that comparison: it supplies the evidence for choosing among candidates, correcting or refining them, and deciding when further investment in measurement is warranted. 

Validation does not privilege (fallible) humans over (fallible) AI. Rather, it privileges explicit, well-defined criteria as a way to make predictions from black-box models more interpretable. 
Validation evidence may come from human-coded labels, scientific instruments, administrative data, or theoretical predictions. 
Human codings are useful insofar as they are auditable implementations of a criterion. 

Ambiguous edge cases and residual disagreement between annotations, while common, do not imply that validation lacks value. 
Instead, they identify where refinement may be needed: the construct definition may be ambiguous, the measurement procedure may be underspecified, or the validation criterion may otherwise require clarification.\footnote{Recent work builds AI-assisted pipelines for surfacing such edge cases \citep[][]{Xiong2025CoDetect} and prompting the researcher to adjudicate them---another example illustrating how AI can transform the construct definition stage of measurement.}
A black-box neural network will often encode richer information than any given criterion. The question is whether this additional variation suggests a better construct, or instead reflects irrelevant features, or patterns that are difficult to interpret.

\subsection{Which Model? Which Prompt?}\label{sec:which-model}

It is useful to make the decision of whether and how to validate AI measures more concrete. 
Consider a researcher with a corpus of documents and a concept they would like to measure. The researcher constructs a careful rubric for the concept, converts it into a prompt, queries a frontier language model, and obtains a label for every document in the corpus at minimal cost. Rather than training annotators to score economic policy uncertainty in newspaper articles \citep{baker2016measuring} or the tone of congressional speeches about immigration \citep{Card2022PoliticalSpeeches}, the researcher simply asks the model. The temptation is to treat these outputs as the (noiseless) measurements themselves or proxies, and proceed directly to the downstream analysis. Is this a credible exercise?

A core challenge that such an approach must confront is that a well-defined construct alone does not pin down the measurement. The researcher must also specify a model, a prompt (when generative AI is used), and various model settings (that are often set by default but can sometimes be changed)---such as the temperature, decoding algorithm (\textit{e.g.,} greedy decoding), and system prompt. The prompt engineering strategy determines how the rubric is converted into instructions for the model: as a system prompt or a user prompt; in full or in summary; with or without worked examples. The researcher may even ask the model to adopt a persona, such as an expert annotator. Each combination of choices produces an alternative measurement of the same concept, and generating yet another alternative is nearly free. Alternatively, if the researcher is not satisfied with off-the-shelf models, they may choose to train their own model, choosing an architecture, training data distribution, and a variety of other important details like the hyperparameters, the tokenizer, and pre- and post-processing steps. Such decisions are free parameters of the measurement process: nothing in the construct itself pins down the choice between AI models or prompt strategies.

These choices can matter enormously for downstream estimates.
Varying the language model and the prompt engineering strategy can move the resulting estimates in magnitude, in statistical significance, and even in sign. 
\citet{LudwigMullainathanRambachan2026} document this sensitivity for measurements of congressional bills and financial news headlines, and a growing literature documents the same phenomenon across annotation tasks in the social sciences \citep{Sclar2024PromptFormatting, Atreja2025WhatsPrompt, Yin2026ExposureScores}. 
\citet{baumann2025large} replicate annotation tasks from 21 published studies and show that, with a handful of prompt paraphrases, virtually any hypothesis can be made to appear statistically significant.\footnote{Recent work builds structured prompting pipelines that make outputs more reliable and reproducible across semantically similar prompts \citep{Asirvatham2026GPTMeasurement}. This work is valuable and can potentially narrow the arbitrariness of these choices.}

In the face of this evidence, one response simply accepts the sensitivity. If different protocols produce different measurements, perhaps we should define the concept to be the output of one particular model under one particular prompt and collection of model settings. This position is internally consistent, but its implications are hard to accept. Are researchers using ChatGPT, Claude, DeepSeek, or any other of a multitude of commercial, open-source, or customized models to measure economic policy uncertainty really studying different constructs? What about two researchers who use the same model but different prompts? And what are we to do when a new frontier model is released? Should we revisit every empirical finding based on the previous generation's outputs? Defining the construct as the output of a specific AI model makes results incomparable across studies, even studies asking the same economic question. More fundamentally, the problem with this view is that measurement should be anchored in the construct, not the other way around.

A second response defends the choices by checking agreement. If several models or prompt strategies tend to assign similar labels, the reasoning goes, then the free parameters must not matter very much. However, agreement among AI-generated measurements is weaker evidence than it appears. Frontier models are built on similar architectures, trained on overlapping corpora, and tuned toward similar preferred answers. An emerging literature on response homogenization documents exactly this convergence \citep{kirk2024understanding, jiang2026artificial}, arguing that pre-training makes models broadly competent in similar ways, and reinforcement learning with human feedback (used to develop frontier models) pushes them toward similar preferred answers, with output variation driven to a considerable degree by prompting. In other words, there are strong theoretical reasons to expect model errors to be correlated---and they are in practice, as documented on the benchmark datasets used to evaluate frontier models \citep{Kim2025CorrelatedErrors}. 

In short, agreement across AI-generated measurements can simply reflect shared architectures and underlying training data, rather than convergence to implementing the construct as the researcher intends. At the same time, because the space of implementation choices is high-dimensional (the prompt space, for instance, is effectively infinite), there are also many levers that when varied can produce diverging responses for the same construct. 

Where does this leave the researcher? The choice among models, prompts, and implementation details is consequential, and no particular choice has a privileged claim to the construct. The researcher can therefore search across models and prompts, deliberately or not, until the downstream estimate looks right: p-hacking in the form of measurement choice. Empirical economics has confronted a version of this problem before. When computation became cheap, researchers could search across regression specifications until a result emerged \citep[e.g.,][]{sala1997just}. The response was to build causal inference toolkits that guide and discipline the researcher's choice of specification. We are now building toolkits that provide the same discipline for AI-generated measurements.

\subsection{Routes to Credible Measurement}\label{sec:routes}

Consider again the researcher who converts a rubric into a prompt and asks an LLM to label every document in the corpus. Section~\ref{sec:which-model} ruled out two easy answers: treating the labels as the measurements themselves, and defending them simply with a few sensitivity checks across alternative AI-generated measures. Two routes remain. The first treats the model as a black box and establishes credibility by validating its outputs. The second opens the black box far enough to justify the assumptions about model behavior that identification requires.

There is a strong case for pursuing the first approach when it is feasible, because the errors made by deep neural networks can shift in potentially complex and meaningful ways with the distribution of inputs and implementation details. This can be difficult to model. Suppose our hypothetical researcher drew a random sample of documents from the corpus and carefully labeled each document according to their rubric.
On this validation sample, the researcher observes both the measurement that aligns with the desired implementation of the rubric and the AI-generated measurement, and so can learn how the model errs relative to their intended construct and correct the downstream estimates accordingly. 

This approach requires no assumptions about how the specific AI model produces its outputs, and that is a central virtue: the model has millions to billions of parameters that were learned in a complex way on data the researcher has never seen. The AI-generated measurements need not be unbiased, or even accurate (although inaccurate measurements will lead to wider confidence intervals).
Instead, their errors are estimated from a random validation sample and corrected.
This idea is an old one in the measurement error literature \citep{Bound1991MeasurementError, LeeSepanski1995, Chen2005MeasurementError, Chen2011NonlinearMeasurementError}, and it has recently been revived for AI-generated measurements \citep{Angelopoulos2023PredictionPowered, EgamiEtAl2024, carlson2025unifying, LudwigMullainathanRambachan2026}.

Using a validation sample in this manner also clarifies the researcher's purpose for using AI. 
The model no longer defines the measurement; it scales an existing one. 
The measurement process the researcher would defend (for example, a documented rubric with adjudicated edge cases applied by trained experts) cannot reasonably cover a massive corpus of documents.
Within this framework, the existing measurement only needs to cover an affordable, high-quality validation sample, and the AI scales it across the rest of the corpus, augmenting researcher-implemented measurements at their best. 
When regressions became cheap, the limiting factor became credible identification strategies \citep{Angrist2010CredibilityRevolution}.
Now that measurement costs are falling, the limiting factor is increasingly having a high-quality measurement process in the first place.
Section~\ref{obs:mar} provides an introduction to this framework.

This is not, however, the only route. 
In some settings, the researcher may not wish to privilege a validation sample, instead viewing all measurements as noisy attempts to capture an underlying latent concept. 
In others, collecting high-quality measurements in the population under analysis is prohibitively costly. 
Validation data may exist, but for a different study, region, or period. 
In both cases, identification rests on assumptions about the behavior of the AI models themselves, and those assumptions bring both new challenges and tools. 
Section~\ref{sec:noval} examines these settings. 

Finally, there are cases where collecting random validation data is impossible, and the conditions required to achieve identification under assumptions about AI model behavior fail. There may still be strong reasons for using AI measurements to study a fundamentally important question and other avenues may exist for suggestive validation. However, it is necessary to proceed with an awareness of the potential pitfalls elaborated above.

\section{Observation with Validation Samples: A Missing-Data Perspective}\label{obs:mar}

\subsection{Inference with AI Predictions as Inference with Missing Structured Data}

We now develop the validation route introduced above. 
We would like a broadly applicable framework that corrects the potential biases introduced by AI measurement, yielding estimates of target parameters that are robust to the choice of measurement instrument (\textit{e.g.,} different models, different prompts, different implementation details). 
At the same time, better measurements should improve precision, thereby clarifying the cost-benefit tradeoff associated with investments in measurement quality.

Foundational work on semiparametric inference with missing data, and in particular the \citet{rubin1976inference} missing-at-random (MAR) mechanism, provides just such a framework. 
It may at first seem surprising to frame inference with AI predictions as a missing-data problem, but MAR is highly suitable because researchers often lack the low-dimensional summaries of unstructured data that are needed for statistical analysis. 
For instance, a researcher may have newspaper articles and use an LLM to impute whether these articles discuss economic policy uncertainty, the variable of interest. 
Because the articles do not include these binary summaries, the problem is fundamentally one of missing data, which the researcher addresses by using an LLM as a scalable imputation technology.  

Rubin's seminal work is closely connected to large, subsequent literatures on semiparametric inference with missing data, measurement error, causal inference (a closely related missing data problem where counterfactual outcomes are missing), and inference with black-box AI predictions.  
Semiparametric missing-data frameworks are particularly complementary to deep neural networks because they allow the data to speak as much as possible, placing minimal assumptions on the deep neural network.
This review follows the treatment in \citet{carlson2025unifying}, which applies Rubin's MAR framework to a variety of settings that are common in inference with AI predictions in economics. 
Readers are referred to this article for elaboration on how their framework, referred to as MAR-S (for \textbf{M}issing \textbf{A}t \textbf{R}andom \textbf{S}tructured Data), relates to the aforementioned literatures.  

The core idea of missing-data frameworks is to use a validation sample to estimate the difference between imputed data and high-quality labels, adjusting target estimates accordingly. Validation data derive from an implementable, well-specified construct definition stated by the researcher. They are obtained through non-scalable, methodically applied processes, such as expert annotation or measurement by scientific ground instruments. The construct may be a highly flawed attempt to capture the concept of interest, but at the observation stage, it is taken as given. 

The central assumption in the MAR-S framework---as the name underscores---is the missing at random assumption: after adjusting for observables, annotated and unannotated observations are comparable in their ground truth values. In other words, there are no unaccounted confounders determining whether an instance of unstructured data is in the validation sample.

\subsection{The Formal Framework}

MAR-S recasts robust and efficient inference with high-dimensional, unstructured data as inference on missing low-dimensional structured data. Structured data, \(M \in \mathcal{M}\), are low-dimensional variables that can be used directly in estimating equations, whereas unstructured data, \(U \in \mathcal{U}\), are high-dimensional and generally unsuitable for direct use in estimation.

Low-dimensional structured data are observed through ``annotation.'' Because this process is too expensive to scale to the full dataset, the researcher estimates an imputation function, \(\hat{\mu}\), to impute the missing structured data. This allows the researcher to leverage the full unstructured dataset, which is often orders of magnitude larger than the validation sample. Increasingly, deep neural networks serve as this imputation function.

Rubin's missing at random mechanism is closely linked to the Rubin Causal Model \citep{rubin1974, imbens2015causal}, and accordingly we use potential outcomes notation. We observe a random variable \(M \in \mathcal{M} \subseteq \mathbb{R}\) that is subject to missingness, with annotation status indicated by \(A \in \{0,1\}\). We write
\[
M = A M^{a=1} + (1-A)M^{a=0}.
\]
We set \(M^{a=0}=0\) with probability one and, for notational convenience, define \(M^* := M^{a=1}\). It follows that
\[
M = A M^*.
\]
We refer to \(M^*\) as the ``ground truth'' potential outcome and to \(A\) as the ``annotation indicator,'' which indicates whether an instance is in the validation sample. We use the term ground truth because it is overwhelmingly used in the literature to denote values in validation samples. To re-emphasize, this does not mean fundamental truth but simply the measurement that derives from careful implementation of a well-specified construct.   

The first assumption is consistency of potential outcomes, which requires annotation status to be well defined and the label for any given instance to depend only on its own annotation status rather than on the annotation status of other instances. In practice, this means the researcher should apply the same measurement criteria, in the same way, to all data instances in the validation sample. 

The central identifying assumption is missing at random. For ground truth potential outcome \(M^* \in \mathcal{M}\), annotation indicator \(A \in \{0,1\}\), observed covariates \(X \in \mathcal{X}\), and unstructured data \(U \in \mathcal{U}\), we assume
\[
(U,M^*) \perp\!\!\!\perp A \mid X.
\]
After adjusting for observables \(X\), annotated and unannotated data are comparable in their ground truth values. This assumption is analogous to ``selection on observables'' in causal inference, a closely related missing data problem where counterfactual outcomes are missing. 

We define the ``annotation score function'' as
\[
\pi(x) := P(A=1 \mid X=x).
\]
The baseline assumption is that \(\pi(x)\) is fixed, known, and bounded away from zero. This embeds the ``strong overlap'' assumption commonly imposed in observational causal inference settings. 

The annotation score will generally be known when the researcher designs the annotation process---and should be documented when data are made publicly available for download by others---but this assumption can be relaxed when this is not the case.

The strong-overlap condition can also be relaxed to settings with decaying overlap. This scenario is relevant to massive data, where it is impossible for the researcher to validate more than a tiny share of a dataset that may contain millions or even billions of observations. As \citet{Kallus2025Surrogates} make clear in their analysis of treatment effects with surrogate outcomes---a problem that parallels AI inference in its underlying structure---there is no need to change the MAR-S estimator itself under an asymptotic regime in which the number of annotations diverges while the ratio of labeled to unlabeled observations converges to zero. The appropriate asymptotic efficiency analysis shows that the variance of the estimator is driven primarily by the size of the labeled dataset rather than by the size of the full dataset.

Additionally, efficiency requires mean squared error consistency:
\[
E\left[\left( \hat{\mu}(\tilde{X})-\mu(\tilde{X}) \right)^2 \right] = o(1).
\]
Intuitively, the expected squared error of the imputation estimator should converge to zero as the amount of data used to train it grows. This condition is needed only for efficiency. It is \textit{not} necessary for unbiased inference. It is relatively mild in the context of deep neural networks.\footnote{If the annotation score function must itself be estimated, the researcher must then assume that the imputation and annotation score functions converge at sufficiently fast rates \citep{kennedy_semiparametric_2023}, for example at \(n^{-1/4}\)-type rates. While convergence rates are unknown given their massive-scale pre-training, these networks appear to converge quickly when being tuned to perform a particular well-defined task.}

\citet{carlson2025unifying} apply this framework to derive estimators for a variety of common empirical scenarios in economics---e.g., ordinary least squares, instrumental variables, difference-in-differences---when left-hand side variables, right-hand side variables, or both are imputed with AI. A package is available for implementing these debiased estimators.\footnote{\url{https://github.com/jscarlson/mar-s}} 

The intuition for MAR-S can be illustrated by deriving the debiased estimator for a mean. Following \citet{Chen2008AuxiliaryData}, who provide general results on semiparametrically efficient estimation of parameters identified by moment conditions with missing data, the moment function for the MAR-S debiased mean is simply $M^*-\theta$. The corresponding estimator is
\[
\hat{\theta}
=
\frac{1}{|\mathcal{I}|}
\sum_{i\in\mathcal{I}}
\left(
\frac{A_i}{\pi(X_i)}
\left[
M_i-\hat{\mu}(\tilde{X}_i)
\right]
+
\hat{\mu}(\tilde{X}_i)
\right),
\]
where \(\mathcal{I}\) is the set of indices allocated to the estimation partition and \(\hat{\mu}\) is an approximation to \(\mu\) learned independently of the estimation sample, either because it is fixed in advance (\textit{e.g.,} by an off-the-shelf model) or because it is fit on a separate tuning split. This is the augmented inverse probability weighted (AIPW) estimator, with \(M^*\) playing the role of a potential outcome.

To better understand the intuition, suppose that
\[
\pi(X_i)
=
\frac{|\mathcal{I}\cap\mathcal{J}|}{|\mathcal{I}|},
\]
where \(\mathcal{J}\) is the set of indices corresponding to annotated observations (in other words, a given share of the instances are randomly annotated). Then the estimator can be written as
\[
\hat{\theta}
=
\underbrace{
\frac{1}{|\mathcal{I}|}
\sum_{i\in\mathcal{I}}
\hat{\mu}(\tilde{X}_i)
}_{A}
+
\underbrace{
\frac{1}{|\mathcal{I}\cap\mathcal{J}|}
\sum_{i\in\mathcal{I}\cap\mathcal{J}}
\left(
M_i^*-\hat{\mu}(\tilde{X}_i)
\right)
}_{B}.
\]
Term \(A\) is the imputation-based estimate of \(E[M_i^*]\) in the estimation sample---what we would get if we ignored the potential measurement error in the neural network predictions---while term \(B\) is a bias-correction term that estimates the measurement error of the imputation function in the annotated sample. This expression is reproduced in recent work on ``prediction-powered inference'' \citep{Angelopoulos2023PredictionPowered}. 
In a big data world, the variance of term A will be small, but the variance of term B could be quite large if AI predictions deviate substantially from $M^*$: noisier predictions will lead to noisier estimates of the target parameter. 

Other estimators are largely analogous to the mean case, whether the imputed variable is an outcome, an instrument, a conditioning variable, etc., and interested readers are referred to \citet{carlson2025unifying} for derivations. 
Two takeaways follow from the general treatment. First, estimators derived under MAR-S, and under related missing-data frameworks, are Neyman orthogonal: they are insensitive to first-order errors in the imputation function, so an inaccurate $\hat{\mu}$ does not compromise the validity of the estimate.  Second, accuracy still matters, but for precision rather than validity. The variance of the estimator falls as $\hat{\mu}$ improves, so the imputation function should be learned as accurately as possible on a held-out sample. In other words, AI predictions, even arbitrarily bad ones, do not endanger a properly debiased estimate; the more accurate the predictor, the more precise the conclusions that can be drawn.

\subsection{Aggregating AI Predictions}

A key limitation of debiasing frameworks is that they assume that validation data are available for the imputed variables used in the estimating equation. In many applications in applied economics, however, validation data are available only at the granular level of individual texts or images, whereas the variable of interest is a potentially nonlinear function or functional of the granular missing data. 

To make debiasing practically useful for many economic applications involving aggregated and transformed AI predictions, \citet{carlson2025unifying} address this mismatch between the level at which ground truth is observed and the level at which the variable enters the estimating equation.

Fortunately, the solution is straightforward to implement in many scenarios. Consider the linear model
\[
Y_i = X^*_i \beta + \varepsilon.
\]
where \(X_i^*\) is an aggregate that is not actually observed, but is instead a function or functional of AI predictions. For example, $X_i^*$ might be mean economic policy uncertainty in year $i$, derived from predictions at the level of individual news articles that are aggregated to the annual level. 

Crucially, valid debiasing---while it does not remove measurement error---transforms any systematic biases from the AI predictions into classical measurement error. Hence, the approach is to first construct a debiased aggregate using MAR-S, and then proceed with the standard method-of-moments correction for classical errors-in-variables \citep{deaton_panel_1985, fuller_measurement_1987}.

This setup extends readily to a number of common empirical settings, including clustering and panel data \citep{deaton_panel_1985}. It can also accommodate heterogeneity in the variance across aggregated observations, cases in which the outcome is itself generated by a MAR-S first step, and non-normal measurement error distributions. \citet{carlson2025unifying} apply it to an analysis from \citet{baker2016measuring} where predictions of whether individual news articles discuss economic policy uncertainty are aggregated to the national annual level, logged, differenced, and interacted with a firm-level characteristic. Classical measurement error frameworks can accommodate these transformations. Importantly, using MAR-S to construct a debiased annual mean, rather than simply treating errors from the neural network as classical, is crucial for making the classical measurement error model plausible.

\subsection{Annotation in Practice}

In the most straightforward cases, researchers can annotate or access validation data for a simple random sample of data instances. However, in other cases, a simple random sample is unlikely to be informative. In large datasets, the structured data of interest may correspond to a rare event. For instance, a researcher may be interested in social media posts about inflation, which are a very small share of the total corpus of social media posts. In the machine learning literature, this is referred to as ``class-imbalanced data.'' 

In such a setting, social scientists often use keyword-based filtering to select texts for annotation: texts containing specific keywords receive some positive probability of annotation, while all other texts are excluded. This approach violates strong overlap when structured data are then imputed from the full corpus, because it assigns zero annotation probability to some data instances. Intuitively, a language model's measurement error is plausibly systematically correlated with the terms that appear in the text, and so annotating only articles with particular terms may not be informative about how the model performs on a broader variety of texts.

A researcher could limit attention to only texts with certain keywords, but in a variety of cases this risks overlooking many relevant texts, leading to a form of selection bias. Fortunately, there are alternatives to keywords for annotating class-imbalanced data. Existing work in machine learning and statistics, which can be incorporated into the MAR-S framework, suggests that researchers should oversample instances that are harder to impute \citep{zrnic_active_2024}.
This aligns closely with a broad literature on importance sampling \citep{mcbook}. 
But this prescription cannot be implemented directly because the difficulty of imputation depends on the unknown ground truth. However, the machine learning literature suggests various observable proxies for imputation difficulty, including model-based variance estimates, cross-validation residuals, and disagreement across ensemble predictions. Another approach that can work well in practice is to calculate the distance in embedding space to some relevant query---\textit{e.g.,} this post is about inflation---where this embedding distance serves as the observable $X$ that determines annotation probability in the MAR-S framework \citep[see][for an elaboration on embedding models and distances]{Dell2025}.

Another frequent question is how large the validation sample should be. The number of effective observations represented by the combined validation and imputed data depends on the asymptotic correlation between the validation sample-only estimator and the imputation-only estimator. For example, in the case of mean estimation, \citet{broska_mixed_2025} define the effective number of observations as
\[
n_0 := |\mathcal{I} \cap \mathcal{J}| \times \frac{|\mathcal{I}|}{|\mathcal{I} \cap \mathcal{J}|+|\mathcal{I} \cap \mathcal{J}^c|\left(1-\max\{\tilde{\rho}^2, 0\}\right)},
\]
where \(\mathcal{J}\) is the set of annotated indices, \(\mathcal{J}^c\) is the set of unannotated indices, and \(\tilde{\rho}\) is the asymptotic correlation between the ground truth only estimator and the imputation-only estimator,
\[
\tilde{\rho}:=\text{Cor}(M^*, \mu(\tilde X)).
\]
The more predictive the imputation function, the greater the effective sample size. In the limiting case of a perfect imputation function, the effective sample size equals the full size of the dataset.

The asymptotic correlation between the ground truth-only estimator and the imputation-only estimator is unknown, so this cannot be calculated directly. Researchers may nevertheless have a prior about it, or wish to calculate effective sample sizes under different scenarios. For a binary classifier (e.g., ``is this post about inflation?''), in practice often a couple hundred high-quality annotations are sufficient when the imputation function is reasonably accurate. As predictions become noisier, incorporating them contributes less and less relative to simply using the small validation sample. In this scenario, it is necessary for the researcher to invest in more accurate AI predictions (by training a custom model, using higher quality training data or a larger model, etc.) if they wish to leverage AI predictions to improve power. 

What about errors in the validation sample? One cannot correct arbitrary measurement error in AI predictions using validation data measured with arbitrary error, although it is possible to make some progress by putting structure on the errors. 
To minimize errors, validation data should be the researcher's measurement at its best: a modestly sized sample built using a carefully defined construct by a process that is tightly supervised by the researcher. AI can scale the measure, so large samples are not needed. 
Rather, the job of the researcher is to ensure the validation data are of the highest quality. If collecting high-quality validation data is not feasible, a different approach to measurement with AI predictions is required. 

\section{Observation without Random Validation Samples}\label{sec:noval}

In Section~\ref{obs:mar}, we saw that the missing at random assumption is analogous to the selection-on-observables assumption in causal inference, which can likewise be viewed as a missing data problem: treatment assignment reveals which potential outcome is observed, whereas in our setting annotation determines whether $M^*$ is observed. Hence, much as randomized treatment is the gold standard for causal inference, random validation samples are the gold standard for inference with unstructured data. Yet, it is often impossible to create a high-quality, random validation sample, much as randomized treatment is often infeasible in causal inference. Fortunately, just as causal inference can proceed under additional assumptions when randomized treatment assignment is infeasible, valid inference with unstructured data can proceed under additional assumptions when a random validation sample cannot be constructed.

The validation sample framework in Section~\ref{obs:mar} rests on two requirements. First, the researcher must specify a measurement process that they are prepared to defend: an explicit rubric or instrument-based measurement that defines the construct $M^*$ and adjudicates edge cases. Second, the researcher must apply that measurement process to a random validation sample drawn from the population under analysis. This section considers scenarios in which each requirement fails.

Sometimes there are no existing validation data, and the researcher cannot create them if no technology can access $M^*$---that is, all measurement technologies are inherently noisy---or applying a technology that can is infeasible (Section~\ref{sec:multiple-measurements}). 
Other times validation data exist or can be compiled, but from another region, another period, or another population, so the labels are not a random validation sample for the analysis at hand (Section~\ref{sec:data-combination}). 
In both cases, identification of the downstream parameter requires additional measurement assumptions, and those assumptions are claims about the behavior of frontier AI and machine learning models.

Why might we be in a setting with non-random or non-existent validation data?
A useful distinction that often matters in practice is ``$U$-measurability.'' 
$M^*$ is $U$-measurable if it can be accessed solely from unstructured data $U$ without consulting any external information. 
For example, simple text analysis tasks are often $U$-measurable. If a construct is $U$-measurable, it will typically be feasible for the researcher to construct a random validation set if they can access $U$ (which may not be the case for externally compiled data). 
There are a wide variety of settings, however, where $M^*$ is not $U$-measurable. This is particularly common in remote sensing. 
An annotator cannot directly observe crop type from lower resolution satellite data, or consumption from nighttime lights. 
Instead, validation requires ground measurements on crop type or consumption that are very costly to collect. 
Often some validation sample exists---without labeled data on crop type or consumption, a model could not have been trained to predict them in the first place. 
But typically the researcher wants to study a different region, time period, or population. 
Sometimes, the entire motivation for using remote sensing is to collect data for areas where obtaining ground measurements is impossible (\textit{e.g.}, remote or conflict prone areas). 
In short, creating a random validation sample for the population of interest is often out-of-reach. 

\subsection{Multiple Measurements}\label{sec:multiple-measurements}

Suppose we are in a world where instead of some non-scalable technology revealing $M^*$, the best each measurement technology $j$ can do is to access $M^* + \eta_j$, where $\eta_j$ is an error term associated with measurement technology $j$. 
The construct is still well-defined, but it can only be applied with noise. (If the construct definition itself is uncertain, a discovery regime, as described in Section~\ref{subsec:disc}, will plausibly be more suitable.) 

In this case, the researcher may instead wish to treat the construct $M^*$ as if it were latent and each measurement protocol were a repeated, noisy measurement of it. 
Concretely, suppose the researcher forms $J$ measurement protocols, producing measurements $\widehat{M}_1, \dots, \widehat{M}_J$ of the same construct.
Latent variables observed only through error-prone measurements are the subject of a rich literature on measurement systems and finite mixtures, stretching from classical models of observer error to modern nonclassical measurement error \citep{Dawid1979ObserverErrorRates, Schennach2020MismeasuredVariables, Schennach2021MeasurementSystems}. 

Might these identification results be useful here? This depends on whether additional assumptions about the AI-generated measurements are met. 

Consider the simple case in which the construct is binary, $M^* \in \{0, 1\}$, and the researcher generates at least three measurements, $J \geq 3$. Classical results dating to \citet{Kruskal1977ThreeWayArrays} and \citet{Dawid1979ObserverErrorRates} show that if the measurement errors are conditionally independent across protocols given the intended construct, and each protocol $j$ is informative (meaning its true positive rate exceeds its false positive rate: $\Pr(\widehat{M}_j = 1 \mid M^* = 1) > \Pr(\widehat{M}_j = 1 \mid M^* = 0)$), then the joint distribution of $(\widehat{M}_1, \dots, \widehat{M}_J)$ identifies both the distribution of the latent construct $M^*$ and each protocol's error rates. 
With enough conditionally independent measurements of the same latent construct, the data themselves reveal how often each protocol errs. No validation sample is required.\footnote{See \citet{Hu2008NonclassicalMeasurement}, \citet[][]{Allman2009Identifiability}, and \citet{Bonhomme2016FiniteMixtures} for generalizations of these ideas to richer constructs, measurements, and error structures.}

When such identification results are applied to AI, the assumptions are not opaque conditions, but rather statements about the behavior of AI models. The informativeness condition is plausibly mild. As long as the researcher is measuring something sufficiently concrete and well-defined, it seems reasonable to assume the AI predictions are better than chance.

The problem is justifying that measurements are conditionally independent across models, prompts, and other implementation details. Conditional independence asserts that different models or different prompts err independently on the same input. There is every structural reason to expect the opposite: frontier models share architectures, overlap heavily in their training corpora, and are tuned with similar human-feedback pipelines \citep{Kleinberg2021AlgorithmicMonoculture, Bommasani2022OutcomeHomogenization}. In some cases, they are directly ``distilled'' from---or trained on data derived from---other frontier models. 

Because conditional independence is a claim about how frontier models err, we can empirically check whether it is true if we are able to access $M^*$. 
As argued in Section~\ref{sec:credible}, a natural place to look is the diverse set of benchmark datasets that are used to evaluate frontier models. Researchers have documented that frontier models err on the same items, and more capable models err more similarly \citep{Kim2025CorrelatedErrors}. The same behavior appears in economic measurement, where errors are strongly correlated across models and prompts \citep{LudwigMullainathanRambachan2026, Chen2026LLMPrompts}. 

The failure of conditional independence does not mean that multiple measurements are uninformative. 
It means the assumption must be relaxed. 
One path allows for bounded violations of conditional independence, with the size of the permitted violation calibrated to external evidence on how correlated model errors actually are in some external dataset(s) for which $M^*$ is available (\textit{e.g.,} AI benchmarks or economic domains where validation data exist). 
The researcher then reports an identified set for the downstream parameter rather than a point estimate, together with a breakdown value: how large would the violation have to be to overturn the conclusion? \citet{Chen2026LLMPrompts} develop this approach for AI-generated measurements, and we refer the interested reader there for further details.
The general lesson is that, given what we know about how these models err, multiple measurements may partially identify downstream parameters, with bounds whose width is governed by that behavior.

To the extent that all measures are generated by a black-box AI model, we might also worry about the construct validity issues raised in Section~\ref{section:pipeline}: how do we know the models are giving multiple measurements of, \textit{e.g.,} trust, rather than something that is simply correlated with it? Here, evidence on convergent and discriminant validity remains useful. 

The measurement error literature offers many identification arguments beyond repeated measurements: for example, based on instrumental variables, auxiliary data, and restrictions on the joint distribution of errors (see \citet{Chen2011NonlinearMeasurementError, Schennach2020MismeasuredVariables, Schennach2021MeasurementSystems} for reviews).
Historically, the assumptions behind these arguments concerned human measurements.
Applied to AI-generated measurements, the same assumptions become properties of models, and models can be queried, tested, and benchmarked. 
Assessing which identification arguments are credible, and how sensitive conclusions are to violations of their assumptions, becomes an empirical exercise: ask what each assumption asserts about the behavior of frontier AI models, and bring evidence to bear on it.

\subsection{Data Combination}\label{sec:data-combination}

We next turn to the second failure of the validation sample framework, in which annotated data exist but not for the population under analysis. Remote sensing is the showcase application. 
Machine learning models applied to satellite imagery and other digital traces now produce measurements of poverty, living standards, deforestation, weather and climate patterns, and crop cover across the globe, including places that may be difficult for survey teams to reach or where ground measurement instruments are difficult to maintain \citep{Hansen2013ForestChange, Blumenstock2015MobilePhonePoverty, Jean2016SatellitePoverty, Rolf2021GlobalSatellite, Aiken2025DigitalTraces}. 
These measurements increasingly serve as outcomes and regressors in development, environmental, and urban economics. 
In these applications, ground truth comes from the traditional measurement processes that the machine learning model seeks to scale: an in-person survey that elicits consumption or income, surveyors sent into forests to measure tree cover, or physical monitors that record pollution. 
When a conditionally random validation sample drawn from the population under analysis exists, the framework of Section~\ref{obs:mar} applies directly \citep{Proctor2023ParameterRecovery, Kluger2025ImputedCovariates, Lu2025RegressionMaps, PELLETIER2026103655}.

Because collecting such data is costly, the machine learning model is commonly trained and validated on an auxiliary sample where ground truth exists, in one country, one period, or one sensor's footprint, and then deployed on the analysis sample, where ground truth has not been collected.
The auxiliary sample is not a validation sample. 
Nothing guarantees that the relationship between predictions and ground truth transfers, and predictive performance often degrades when transferring across regions and periods \citep[e.g.,][]{Proctor2025SatelliteMeasure}. More generally, a central limitation of neural networks is that performance tends to decline with domain shift away from the training data distribution \citep{ben2010theory}. 

Often, validation data do not exist precisely because regions are fundamentally different (\textit{e.g.,} more geographically isolated, poorer, or conflict prone) from areas where ground data collection is feasible. 
These concerns are compounded when, as is often the case, the only available evaluation sample is a held-out split from the same annotated dataset used for model training, since performance on a substantively different target population may shift in complex ways relative to performance on the in-sample distribution.

To proceed, the researcher must assume that some feature of the auxiliary sample carries over to the analysis sample.
If this is not plausible, researchers can consider whether the multiple measurement framework is a better fit. 

The literature has organized around two alternative identifying assumptions for this data combination problem. Let $a$ denote the auxiliary sample, in which both the AI-generated measurement $\widehat{M}$ and the ground truth $M^*$ are observed; let $e$ denote the analysis sample, in which only $\widehat{M}$ is observed; and let $P_a$ and $P_e$ denote the corresponding distributions.

The first identifying assumption, which we call \emph{outcome stability}, states that the distribution of the ground truth given the prediction is the same in both samples: $P_a(M^* \mid \widehat{M}) = P_e(M^* \mid \widehat{M})$. 
In the auxiliary sample, the researcher learns the conditional distribution of the ground truth measurement given the prediction, and this conditional distribution transfers to the analysis sample. 
Outcome stability may be plausible when the prediction was constructed to forecast the outcome, as when a poverty score is built from phone metadata. 
It is the same assumption underlying the literature on ``surrogate outcomes'' \citep{Chen2008AuxiliaryData, Kallus2025Surrogates, Athey2026SurrogateIndex}.

To see what outcome stability delivers, consider estimating the prevalence of a binary construct in the analysis sample, $\theta = P_e(M^* = 1)$. 
Under outcome stability, the researcher can take the conditional distribution learned in the auxiliary sample, apply it to each prediction in the analysis sample, and average: $\theta = \mathbb{E}_e[P_a(M^* = 1 \mid \widehat{M})]$. 
The argument parallels the validation sample analysis of Section~\ref{obs:mar}, in which the researcher also learns the relationship between the ground truth and the AI-generated measurements on an annotated subsample and extrapolates to the rest of the sample. 
The difference is that random annotation guarantees the extrapolation is valid by design, whereas here it must be assumed. 
Making this argument precise requires additional conditions; we refer the reader to existing work for formal statements \citep{Kallus2025Surrogates, Athey2026SurrogateIndex}.

The second identifying assumption, which we call \emph{measurement stability}, states that the distribution of the prediction given the ground truth is the same in both samples: $P_a(\widehat{M} \mid M^*) = P_e(\widehat{M} \mid M^*)$. 
In the auxiliary sample, the researcher learns how the AI-generated measurement errs given the truth, that is, the model's error rates, and these error rates transfer to the analysis sample. 
Measurement stability may be plausible when the outcome physically generates the signal that the model reads, as when crop burning produces the smoke plume that a satellite sensor detects.
\citet{Rambachan2026RemoteOutcomes} study identification under measurement stability for AI-generated measurements in remote sensing.

Under measurement stability, the construct's prevalence is identified by a different argument \citep{Rambachan2026RemoteOutcomes}. The rate of positive predictions in the analysis sample mixes together the model's error rates: $P_e(\widehat{M} = 1) = \theta \, P_a(\widehat{M} = 1 \mid M^* = 1) + (1 - \theta) \, P_a(\widehat{M} = 1 \mid M^* = 0)$. Provided the predictions are informative, in the same sense as in Section~\ref{sec:multiple-measurements}, the researcher can invert this equation, correcting the analysis sample's prediction rate using the auxiliary sample's error rates. Notice that the two stability assumptions use the same data but produce different formulas, and in general they produce different answers.

Which stability assumption should the researcher invoke, if either? 
The substance of the choice is which conditional distribution is plausibly invariant across the two samples. 
Measurement stability asserts that the way the measurement is generated from the underlying construct is stable across the samples, for example, the way crop burning or deforestation appears in satellite images. 
It therefore faces engineering threats, such as changes in sensors, image resolution, or model versions between the samples. 
Outcome stability asserts that the relationship between the construct and the prediction is stable, and so anything that alters the distribution of the construct given the prediction across the samples is a potential threat.
Stating which threats are plausible in the application at hand is the analogue, in this setting, of defending an identifying assumption in a research design. 
The choice is worth making carefully: exploiting the correct stability assumption can deliver substantially more precise estimates than discarding the auxiliary sample or applying the predictions naively \citep{Rambachan2026RemoteOutcomes, Alsharif2026CausalSatellite}.

Like conditional independence in Section~\ref{sec:multiple-measurements}, both measurement stability and outcome stability are assumptions about the behavior of frontier AI and machine learning models, in this case about how that behavior travels across settings. 
These too are empirical claims, and benchmarks are again the natural place to assess them. 
The existing benchmarks, however, measure a different property: they document how predictive performance degrades across regions, periods, and sensors.
But predictive performance does not reveal whether outcome stability or measurement stability holds. 
Building benchmarks that track whether these conditional distributions transfer, on the measurement tasks economists care about, is a natural next step and would put the choice between stability assumptions on an empirical footing.

Measurement without a random validation sample is not measurement without discipline. 
The discipline instead moves from the design of the validation sample to the defense of the assumption. 
Building that discipline, through new identification arguments and new evidence on how models err, is an active area of research.
The researcher's obligations follow: state the assumption explicitly, defend it in the application at hand, evaluate it where evidence exists, and report how conclusions change as it is relaxed. 

\section{Conclusion}\label{sec:checklist}

Empirical economics is limited by what we can measure. Structured data record only features someone chose to observe, while careful, interpretable measurement is often too costly to implement at scale.
AI relaxes both constraints. 
Unstructured data capture dimensions of economic life that have rarely or never been measured systematically, and AI can transform them into structured variables, allowing measurement processes once feasible only on small samples to be applied across entire corpora. 
Opportunities to apply AI arise throughout the measurement pipeline, from discovering candidate constructs to defining them and implementing them at scale. 
AI is therefore widening the aperture of what economics can measure.

This article focuses primarily on observation, where AI usage is by far the most widespread and where the associated statistical challenges are most developed.
The central idea is simple: apply a defensible measurement process to a random validation sample and compare these measurements to the AI generated measures to correct downstream estimates. This logic predates AI: statistical agencies, for example, have long used expensive, high-quality validation samples to correct much cheaper census measurements collected across the full population.

This framework is particularly well suited to black-box AI because researchers need not model the neural network, assume its predictions are unbiased, or even require them to be accurate.
Instead, prediction errors are learned from a validation design the researcher controls and should therefore be able to interpret and defend. 
Even a model with billions of parameters trained on inaccessible data can thus support credible, interpretable empirical estimates.
When random ground truth cannot be collected, however, stronger assumptions become unavoidable. 
This literature is still developing, and an important direction for future work is to connect evaluations of frontier models more directly to the assumptions required for credible empirical research.

In Appendix~\ref{app:checklist}, we provide a checklist for researchers using AI-generated variables. 
It is designed both to help authors apply the framework developed in this article and to guide transparent reporting.
It identifies which of the three cases a study falls into, clarifies the relevant assumptions and requirements, and specifies what should be reported about the measurement process---including construct definition, validation-label generation, and the model and (if applicable) prompt used.
We hope it promotes more rigorous and transparent empirical use of AI-generated measurements.

\clearpage
\begingroup
\bibliographystyle{aea}
\bibliography{cites}

@article{sala1997just,
  title={I Just Ran Two Million Regressions},
  author={Sala-I-Martin, Xavier X},
  journal={The American Economic Review},
  pages={178--183},
  year={1997},
  publisher={JSTOR}
}

@article{cronbach1955construct,
  title={Construct validity in psychological tests.},
  author={Cronbach, Lee J and Meehl, Paul E},
  journal={Psychological Bulletin},
  volume={52},
  number={4},
  pages={281},
  year={1955},
  publisher={American Psychological Association}
}

@misc{BaronEtAl2026,
  author = {Baron, Jason and Dobbie, Will and Lombardo, Richard and Rambachan, Ashesh and Ryan, Joseph},
  title  = {Human Decisions and Machine Predictions in Multistage Systems},
  year   = {2026},
  note   = {Working paper, June 29, 2026}
}

@misc{BatistaRoss2024,
  author = {Batista, Rafael M. and Ross, James},
  title  = {Words that Work: Using Large Language Models to Generate and Refine Hypotheses from Text},
  year   = {2024},
  note   = {Working paper}
}

@article{Carlson2026,
  author        = {Carlson, Jacob},
  title         = {Making Interpretable Discoveries from Unstructured Data: A High-Dimensional Multiple Hypothesis Testing Approach},
  year          = {2026},
  journal={arXiv preprint arXiv:2511.01680},
}

@misc{ChopraHaaland2026,
  author = {Chopra, Felix and Haaland, Ingar},
  title  = {Conducting Qualitative Interviews with {AI}},
  year   = {2026},
  note   = {Working paper, March 9, 2026}
}

@misc{GeieckeJaravel2026,
  author = {Geiecke, Friedrich and Jaravel, Xavier},
  title  = {Conversations at Scale: Robust {AI}-led Interviews},
  year   = {2026},
  note   = {Working paper, February 7, 2026}
}

@inproceedings{VafaEtAl2024,
  author    = {Vafa, Keyon and Rambachan, Ashesh and Mullainathan, Sendhil},
  title     = {Do Large Language Models Perform the Way People Expect? {M}easuring the Human Generalization Function},
  booktitle = {Proceedings of the 41st International Conference on Machine Learning (ICML)},
  year      = {2024}
}

@techreport{HaalandEtAl2024,
  author      = {Haaland, Ingar K. and Roth, Christopher and Stantcheva, Stefanie and Wohlfart, Johannes},
  title       = {Understanding Economic Behavior Using Open-ended Survey Data},
  institution = {National Bureau of Economic Research},
  type        = {NBER Working Paper},
  number      = {32421},
  year        = {2024},
  doi         = {10.3386/w32421}
}

@article{LudwigMullainathan2024,
  author  = {Ludwig, Jens and Mullainathan, Sendhil},
  title   = {Machine Learning as a Tool for Hypothesis Generation},
  journal = {The Quarterly Journal of Economics},
  volume  = {139},
  number  = {2},
  pages   = {751--827},
  year    = {2024},
  doi     = {10.1093/qje/qjad055}
}

@misc{ModarressiEtAl2025,
  author        = {Modarressi, Iman and Spiess, Jann and Venugopal, Amar},
  title         = {Causal Inference on Outcomes Learned from Text},
  year          = {2025},
  eprint        = {2503.00725},
  archivePrefix = {arXiv},
  primaryClass  = {stat.ME}
}

@article{MullainathanSpiess2017,
  author  = {Mullainathan, Sendhil and Spiess, Jann},
  title   = {Machine Learning: An Applied Econometric Approach},
  journal = {Journal of Economic Perspectives},
  volume  = {31},
  number  = {2},
  pages   = {87--106},
  year    = {2017}
}

@inproceedings{MovvaEtAl2025,
  author    = {Movva, Rajiv and Peng, Kenny and Garg, Nikhil and Kleinberg, Jon and Pierson, Emma},
  title     = {Sparse Autoencoders for Hypothesis Generation},
  booktitle = {Proceedings of the 42nd International Conference on Machine Learning (ICML)},
  year      = {2025}
}

@inproceedings{MovvaEtAl2026,
  author    = {Movva, Rajiv and Milli, Smitha and Min, Sewon and Pierson, Emma},
  title     = {What's in My Human Feedback? Learning Interpretable Descriptions of Preference Data},
  booktitle = {International Conference on Learning Representations (ICLR)},
  year      = {2026}
}

@misc{MullainathanRambachanScience2025,
  author = {Mullainathan, Sendhil and Rambachan, Ashesh},
  title  = {Science in the Age of Algorithms},
  year   = {2025},
  note   = {Working paper, October 10, 2025}
}

@article{ObermeyerEtAl2026,
  author  = {Obermeyer, Ziad and Schubert, Alexander and Ross, James and Mullainathan, Sendhil and Lingman, Markus},
  title   = {An ECG Biomarker for Sudden Cardiac Death Discovered with Deep Learning},
  journal = {Nature},
  year    = {2026},
  doi     = {10.1038/s41586-026-10674-6},
  note    = {Published online 24 June 2026}
}

@inproceedings{PengEtAl2025,
  author        = {Peng, Kenny and Movva, Rajiv and Kleinberg, Jon and Pierson, Emma and Garg, Nikhil},
  title         = {Position: Use Sparse Autoencoders to Discover Unknowns},
  booktitle     = {Proceedings of the 43rd International Conference on Machine Learning (ICML)},
  year          = {2026},
  eprint        = {2506.23845},
  archivePrefix = {arXiv},
  primaryClass  = {cs.LG}
}

@book{fuller_measurement_1987,
  author    = {Fuller, Wayne A.},
  title     = {Measurement Error Models},
  publisher = {Wiley},
  address   = {New York},
  year      = {1987}
}

@article{deaton_panel_1985,
	title = {Panel data from time series of cross-sections},
	volume = {30},
	copyright = {https://www.elsevier.com/tdm/userlicense/1.0/},
	issn = {03044076},
	url = {https://linkinghub.elsevier.com/retrieve/pii/0304407685901344},
	doi = {10.1016/0304-4076(85)90134-4},
	language = {en},
	number = {1-2},
	urldate = {2025-02-26},
	journal = {Journal of Econometrics},
	author = {Deaton, Angus},
	month = oct,
	year = {1985},
	pages = {109--126},
}

@techreport{broska_mixed_2025,
  author      = {Broska, David and Howes, Michael and Van Loon, Austin},
  title       = {The Mixed Subjects Design: Treating Large Language Models as Potentially Informative Observations},
  institution = {SSRN},
  type        = {Working Paper},
  number      = {5133034},
  year        = {2025}
}

@article{baumann2025large,
  title={Large language model hacking: Quantifying the hidden risks of using llms for text annotation},
  author={Baumann, Joachim and R{\"o}ttger, Paul and Urman, Aleksandra and Wendsj{\"o}, Albert and Plaza-del-Arco, Flor Miriam and Gruber, Johannes B and Hovy, Dirk},
  journal={arXiv preprint arXiv:2509.08825},
  year={2025}
}

@article{zrnic_active_2024,
	title = {Active {Statistical} {Inference}},
	url = {https://proceedings.mlr.press/v235/zrnic24a.html},
	language = {en},
	urldate = {2024-10-31},
	journal = {Proceedings of the 41st {International} {Conference} on {Machine} {Learning}},
	publisher = {PMLR},
	author = {Zrnic, Tijana and Candès, Emmanuel J.},
	month = jul,
	year = {2024},
	pages = {62993--63010},
}

@book{mcbook,
  author    = {Owen, Art B.},
  title     = {Monte Carlo Theory, Methods and Examples},
  year      = {2013},
  publisher = {\url{https://artowen.su.domains/mc/}}
}

@incollection{Borges1998Exactitude,
  author    = {Borges, Jorge Luis},
  title     = {On Exactitude in Science},
  booktitle = {Collected Fictions},
  translator = {Andrew Hurley},
  year      = {1998},
  publisher = {Viking},
  address   = {New York},
  pages     = {325}
}

@article{ben2010theory,
  title={A theory of learning from different domains},
  author={Ben-David, Shai and Blitzer, John and Crammer, Koby and Kulesza, Alex and Pereira, Fernando and Vaughan, Jennifer Wortman},
  journal={Machine Learning},
  volume={79},
  pages={151--175},
  year={2010},
  publisher={Springer}
}

@inproceedings{kirk2024understanding,
  title={Understanding the effects of rlhf on llm generalisation and diversity},
  author={Kirk, Robert and Mediratta, Ishita and Nalmpantis, Christoforos and Luketina, Jelena and Hambro, Eric and Grefenstette, Edward and Raileanu, Roberta},
  booktitle={International Conference on Learning Representations},
  volume={2024},
  pages={20620--20653},
  year={2024}
}

@article{jiang2026artificial,
  title={Artificial hivemind: The open-ended homogeneity of language models (and beyond)},
  author={Jiang, Liwei and Chai, Yuanjun and Li, Margaret and Liu, Mickel and Fok, Raymond and Dziri, Nouha and Tsvetkov, Yulia and Sap, Maarten and Choi, Yejin},
  journal={Advances in Neural Information Processing Systems},
  volume={38},
  year={2026}
}

@misc{WangEtAl2026,
  author        = {Wang, Jenny S. and Saperstein, Aliya and Pierson, Emma},
  title         = {In Your Own Words: Computationally Identifying Interpretable Themes in Free-Text Survey Data},
  year          = {2026},
  eprint        = {2603.26930},
  archivePrefix = {arXiv},
  primaryClass  = {cs.CY}
}

@misc{Workman2025,
  author = {Workman, Ben},
  title  = {Inside the Black Box: Using Machine Learning to Discover Characteristics of Effective Teaching},
  year   = {2025},
  note   = {Working paper}
}

@inproceedings{ZhongEtAl2022,
  author    = {Zhong, Ruiqi and Snell, Charlie and Klein, Dan and Steinhardt, Jacob},
  title     = {Describing Differences between Text Distributions with Natural Language},
  booktitle = {Proceedings of the 39th International Conference on Machine Learning (ICML)},
  year      = {2022}
}

@inproceedings{ZhongEtAl2023,
  author    = {Zhong, Ruiqi and Zhang, Peter and Li, Steve and Ahn, Jinwoo and Klein, Dan and Steinhardt, Jacob},
  title     = {Goal Driven Discovery of Distributional Differences via Language Descriptions},
  booktitle = {Advances in Neural Information Processing Systems (NeurIPS)},
  year      = {2023}
}

@misc{ZhouEtAl2024,
  author        = {Zhou, Yangqiaoyu and Liu, Haokun and Srivastava, Tejes and Mei, Hongyuan and Tan, Chenhao},
  title         = {Hypothesis Generation with Large Language Models},
  year          = {2024},
  eprint        = {2404.04326},
  archivePrefix = {arXiv},
  primaryClass  = {cs.AI}
}

@inproceedings{ZhuEtAl2026,
  author    = {Zhu, Jian-Qiao and Xie, Hanbo and Arumugam, Dilip and Wilson, Robert C. and Griffiths, Thomas L.},
  title     = {Using Reinforcement Learning to Train Large Language Models to Explain Human Decisions},
  booktitle = {International Conference on Learning Representations (ICLR)},
  year      = {2026}
}

@article{Donoho2024,
  author  = {Donoho, David},
  title   = {Data Science at the Singularity},
  journal = {Harvard Data Science Review},
  volume  = {6},
  number  = {1},
  year    = {2024},
  doi     = {10.1162/99608f92.b91339ef}
}

@article{Dell2025,
  author  = {Dell, Melissa},
  title   = {Deep Learning for Economists},
  journal = {Journal of Economic Literature},
  volume  = {63},
  number  = {1},
  pages   = {5--58},
  year    = {2025}
}

@misc{LiuEtAl2025,
  author        = {Liu, Haokun and Huang, Sicong and Hu, Jingyu and Zhou, Yangqiaoyu and Tan, Chenhao},
  title         = {{HypoBench}: Towards Systematic and Principled Benchmarking for Hypothesis Generation},
  year          = {2025},
  eprint        = {2504.11524},
  archivePrefix = {arXiv},
  primaryClass  = {cs.AI}
}

@article{LudwigMullainathanRambachan2026,
  author  = {Ludwig, Jens and Mullainathan, Sendhil and Rambachan, Ashesh},
  title   = {Large Language Models: An Applied Econometric Framework},
  journal = {Annual Review of Economics},
  year    = {2026},
  doi     = {10.1146/annurev-economics-120925-105620},
  url     = {https://doi.org/10.1146/annurev-economics-120925-105620}
}

@misc{EgamiEtAl2024,
  author = {Egami, Naoki and Hinck, Musashi and Stewart, Brandon M. and Wei, Hanying},
  title  = {Using Large Language Model Annotations for the Social Sciences: A General Framework of Using Predicted Variables in Downstream Analyses},
  year   = {2024},
  url    = {https://naokiegami.com/paper/dsl_ss.pdf},
  note   = {Working paper, November 17, 2024}
}

@article{baker2016measuring,
  title={Measuring economic policy uncertainty},
  author={Baker, Scott R and Bloom, Nicholas and Davis, Steven J},
  journal={The Quarterly Journal of Economics},
  volume={131},
  number={4},
  pages={1593--1636},
  year={2016},
  publisher={Oxford University Press}
}

@article{Bound1991MeasurementError,
    author  = {Bound, John and Krueger, Alan B.},
    title   = {The Extent of Measurement Error in Longitudinal Earnings Data: Do Two Wrongs Make a Right?},
    journal = {Journal of Labor Economics},
    year    = {1991},
    volume  = {9},
    number  = {1},
    pages   = {1--24}
}

@article{Chen2005MeasurementError,
    author  = {Chen, Xiaohong and Hong, Han and Tamer, Elie},
    title   = {Measurement Error Models with Auxiliary Data},
    journal = {The Review of Economic Studies},
    year    = {2005},
    volume  = {72},
    number  = {2},
    pages   = {343--366}
}

@article{Chen2011NonlinearMeasurementError,
    author  = {Chen, Xiaohong and Hong, Han and Nekipelov, Denis},
    title   = {Nonlinear Models of Measurement Errors},
    journal = {Journal of Economic Literature},
    year    = {2011},
    volume  = {49},
    number  = {4},
    pages   = {901--937}
}

@article{Hu2008NonclassicalMeasurement,
    author  = {Hu, Yingyao and Schennach, Susanne M.},
    title   = {Instrumental Variable Treatment of Nonclassical Measurement Error Models},
    journal = {Econometrica},
    year    = {2008},
    volume  = {76},
    number  = {1},
    pages   = {195--216},
    doi     = {10.1111/j.0012-9682.2008.00823.x}
}

@article{Chen2008AuxiliaryData,
    author  = {Chen, Xiaohong and Hong, Han and Tarozzi, Alessandro},
    title   = {Semiparametric Efficiency in {GMM} Models with Auxiliary Data},
    journal = {The Annals of Statistics},
    year    = {2008},
    volume  = {36},
    number  = {2},
    pages   = {808--843},
    doi     = {10.1214/009053607000000947}
}

@misc{Chen2026LLMPrompts,
    author        = {Chen, Xiaohong and Rambachan, Ashesh and Tamer, Elie},
    title         = {Partial Identification from {LLM} Prompts},
    year          = {2026},
    eprint        = {2606.15031},
    archiveprefix = {arXiv},
    note          = {June 25, 2026 version}
}

@article{Alsharif2026CausalSatellite,
    author  = {Alsharif, Haya and Rambachan, Ashesh and Singh, Rahul and Viviano, Davide},
    title   = {Causal Inference with Satellite Imagery: A Comparison of Methods for Forest Conservation Data},
    journal = {AEA Papers and Proceedings},
    year    = {2026},
    volume  = {116},
    pages   = {87--91},
    doi     = {10.1257/pandp.20261019}
}

@article{Angelopoulos2023PredictionPowered,
    author  = {Angelopoulos, Anastasios N. and Bates, Stephen and Fannjiang, Clara and Jordan, Michael I. and Zrnic, Tijana},
    title   = {Prediction-Powered Inference},
    journal = {Science},
    year    = {2023},
    volume  = {382},
    number  = {6671},
    pages   = {669--674},
    doi     = {10.1126/science.adi6000}
}

@techreport{Asirvatham2026GPTMeasurement,
    author      = {Asirvatham, Hemanth and Mokski, Elliott and Shleifer, Andrei},
    title       = {{GPT} as a Measurement Tool},
    year        = {2026},
    number      = {34834},
    institution = {National Bureau of Economic Research},
    type        = {NBER Working Paper},
    doi         = {10.3386/w34834}
}

@article{LeeSepanski1995,
    author  = {Lee, Lung-fei and Sepanski, Jungsywan H.},
    title   = {Estimation of Linear and Nonlinear Errors-in-Variables Models Using Validation Data},
    journal = {Journal of the American Statistical Association},
    year    = {1995},
    volume  = {90},
    number  = {429},
    pages   = {130--140},
    doi     = {10.1080/01621459.1995.10476495}
}

@inproceedings{Xiong2025CoDetect,
    author    = {Xiong, Chenfei and Ni, Jingwei and Fan, Yu and Zouhar, Vil{\'e}m and Rooein, Donya and Calvo-Bartolom{\'e}, Lorena and Hoyle, Alexander and Jin, Zhijing and Sachan, Mrinmaya and Leippold, Markus and Hovy, Dirk and El-Assady, Mennatallah and Ash, Elliott},
    title     = {{Co-DETECT}: Collaborative Discovery of Edge Cases in Text Classification},
    booktitle = {Proceedings of the 2025 Conference on Empirical Methods in Natural Language Processing: System Demonstrations},
    year      = {2025},
    pages     = {354--364},
    publisher = {Association for Computational Linguistics},
    doi       = {10.18653/v1/2025.emnlp-demos.25}
}

@article{Card2022PoliticalSpeeches,
    author  = {Card, Dallas and Chang, Serina and Becker, Chris and Mendelsohn, Julia and Voigt, Rob and Boustan, Leah and Abramitzky, Ran and Jurafsky, Dan},
    title   = {Computational Analysis of 140 Years of {US} Political Speeches Reveals More Positive but Increasingly Polarized Framing of Immigration},
    journal = {Proceedings of the National Academy of Sciences},
    year    = {2022},
    volume  = {119},
    number  = {31},
    pages   = {e2120510119},
    doi     = {10.1073/pnas.2120510119}
}

@inproceedings{Sclar2024PromptFormatting,
    author    = {Sclar, Melanie and Choi, Yejin and Tsvetkov, Yulia and Suhr, Alane},
    title     = {Quantifying Language Models' Sensitivity to Spurious Features in Prompt Design or: How I Learned to Start Worrying about Prompt Formatting},
    booktitle = {International Conference on Learning Representations},
    year      = {2024}
}

@article{Atreja2025WhatsPrompt,
    author    = {Atreja, Shubham and Ashkinaze, Joshua and Li, Lingyao and Mendelsohn, Julia and Hemphill, Libby},
    title     = {What's in a Prompt? A Large-Scale Experiment to Assess the Impact of Prompt Design on the Compliance and Accuracy of {LLM}-Generated Text Annotations},
    journal = {Proceedings of the Nineteenth International AAAI Conference on Web and Social Media},
    year      = {2025},
    pages     = {122--145},
    doi       = {10.1609/icwsm.v19i1.35807}
}

@techreport{Yin2026ExposureScores,
    author      = {Yin, Michelle and Vu, Hoa and Persico, Claudia},
    title       = {How (un)Stable Are {LLM} Occupational Exposure Scores? Evidence from Multi-Model Replication},
    year        = {2026},
    number      = {35110},
    institution = {National Bureau of Economic Research},
    type        = {NBER Working Paper},
    doi         = {10.3386/w35110}
}

@article{Dawid1979ObserverErrorRates,
    author  = {Dawid, A. P. and Skene, A. M.},
    title   = {Maximum Likelihood Estimation of Observer Error-Rates Using the {EM} Algorithm},
    journal = {Journal of the Royal Statistical Society Series C: Applied Statistics},
    year    = {1979},
    volume  = {28},
    number  = {1},
    pages   = {20--28},
    doi     = {10.2307/2346806}
}

@article{Bonhomme2016FiniteMixtures,
    author  = {Bonhomme, St{\'e}phane and Jochmans, Koen and Robin, Jean-Marc},
    title   = {Non-parametric Estimation of Finite Mixtures from Repeated Measurements},
    journal = {Journal of the Royal Statistical Society Series B: Statistical Methodology},
    year    = {2016},
    volume  = {78},
    number  = {1},
    pages   = {211--229},
    doi     = {10.1111/rssb.12110}
}

@incollection{Schennach2020MismeasuredVariables,
    author    = {Schennach, Susanne M.},
    editor    = {Durlauf, Steven N. and Hansen, Lars Peter and Heckman, James J. and Matzkin, Rosa L.},
    title     = {Mismeasured and Unobserved Variables},
    booktitle = {Handbook of Econometrics},
    year      = {2020},
    volume    = {7A},
    pages     = {487--565},
    publisher = {Elsevier},
    doi       = {10.1016/bs.hoe.2020.07.001}
}

@techreport{Schennach2021MeasurementSystems,
    author      = {Schennach, Susanne M.},
    title       = {Measurement Systems},
    year        = {2021},
    number      = {CWP12/21},
    institution = {Centre for Microdata Methods and Practice},
    type        = {cemmap Working Paper},
    doi         = {10.47004/wp.cem.2021.1221}
}

@article{Kleinberg2021AlgorithmicMonoculture,
    author  = {Kleinberg, Jon and Raghavan, Manish},
    title   = {Algorithmic Monoculture and Social Welfare},
    journal = {Proceedings of the National Academy of Sciences},
    year    = {2021},
    volume  = {118},
    number  = {22},
    pages   = {e2018340118},
    doi     = {10.1073/pnas.2018340118}
}

@inproceedings{Bommasani2022OutcomeHomogenization,
    author    = {Bommasani, Rishi and Creel, Kathleen A. and Kumar, Ananya and Jurafsky, Dan and Liang, Percy},
    title     = {Picking on the Same Person: Does Algorithmic Monoculture Lead to Outcome Homogenization?},
    booktitle = {Advances in Neural Information Processing Systems},
    year      = {2022},
    volume    = {35},
    url       = {https://proceedings.neurips.cc/paper_files/paper/2022/hash/17a234c91f746d9625a75cf8a8731ee2-Abstract-Conference.html}
}

@inproceedings{Kim2025CorrelatedErrors,
    author    = {Kim, Elliot Myunghoon and Garg, Avi and Peng, Kenny and Garg, Nikhil},
    title     = {Correlated Errors in Large Language Models},
    booktitle = {Proceedings of the 42nd International Conference on Machine Learning},
    year      = {2025},
    volume    = {267},
    pages     = {30038--30066},
    publisher = {PMLR},
    series    = {Proceedings of Machine Learning Research}
}

@misc{Kluger2025ImputedCovariates,
    author        = {Kluger, Dan M. and Lu, Kerri and Zrnic, Tijana and Wang, Sherrie and Bates, Stephen},
    title         = {Prediction-Powered Inference with Imputed Covariates and Nonuniform Sampling},
    year          = {2025},
    eprint        = {2501.18577},
    archiveprefix = {arXiv}
}

@article{Angrist2010CredibilityRevolution,
    author  = {Angrist, Joshua D. and Pischke, J{\"o}rn-Steffen},
    title   = {The Credibility Revolution in Empirical Economics: How Better Research Design Is Taking the Con out of Econometrics},
    journal = {Journal of Economic Perspectives},
    year    = {2010},
    volume  = {24},
    number  = {2},
    pages   = {3--30},
    doi     = {10.1257/jep.24.2.3}
}

@article{Blumenstock2015MobilePhonePoverty,
    author  = {Blumenstock, Joshua E. and Cadamuro, Gabriel and On, Robert},
    title   = {Predicting Poverty and Wealth from Mobile Phone Metadata},
    journal = {Science},
    year    = {2015},
    volume  = {350},
    number  = {6264},
    pages   = {1073--1076},
    doi     = {10.1126/science.aac4420}
}

@article{Jean2016SatellitePoverty,
    author  = {Jean, Neal and Burke, Marshall and Xie, Michael and Davis, W. Matthew Alampay and Lobell, David B. and Ermon, Stefano},
    title   = {Combining Satellite Imagery and Machine Learning to Predict Poverty},
    journal = {Science},
    year    = {2016},
    volume  = {353},
    number  = {6301},
    pages   = {790--794},
    doi     = {10.1126/science.aaf7894}
}

@techreport{Proctor2023ParameterRecovery,
    author      = {Proctor, Jonathan and Carleton, Tamma and Sum, Sandy},
    title       = {Parameter Recovery Using Remotely Sensed Variables},
    year        = {2023},
    number      = {30861},
    institution = {National Bureau of Economic Research},
    type        = {NBER Working Paper},
    doi         = {10.3386/w30861}
}

@techreport{Proctor2025SatelliteMeasure,
    author      = {Proctor, Jonathan and Carleton, Tamma and Chong, Trinetta and Fransen, Taryn and Greenhill, Simon and Katz, Jessica and Murayama, Hikari and Sherman, Luke and Tseng, Jeanette and Druckenmiller, Hannah and Hsiang, Solomon},
    title       = {What Can Satellite Imagery and Machine Learning Measure?},
    year        = {2025},
    number      = {34315},
    institution = {National Bureau of Economic Research},
    type        = {NBER Working Paper},
    doi         = {10.3386/w34315}
}

@misc{Rambachan2026RemoteOutcomes,
    author        = {Rambachan, Ashesh and Singh, Rahul and Viviano, Davide},
    title         = {Program Evaluation with Remotely Sensed Outcomes},
    year          = {2026},
    eprint        = {2411.10959},
    archiveprefix = {arXiv}
}

@article{Rolf2021GlobalSatellite,
    author  = {Rolf, Esther and Proctor, Jonathan and Carleton, Tamma and Bolliger, Ian and Shankar, Vaishaal and Ishihara, Miyabi and Recht, Benjamin and Hsiang, Solomon},
    title   = {A Generalizable and Accessible Approach to Machine Learning with Global Satellite Imagery},
    journal = {Nature Communications},
    year    = {2021},
    volume  = {12},
    pages   = {4392},
    doi     = {10.1038/s41467-021-24638-z}
}

@article{Aiken2025DigitalTraces,
    author  = {Aiken, Emily and Bellue, Suzanne and Blumenstock, Joshua E. and Karlan, Dean and Udry, Christopher},
    title   = {Estimating Impact with Surveys versus Digital Traces: Evidence from Randomized Cash Transfers in {Togo}},
    journal = {Journal of Development Economics},
    year    = {2025},
    volume  = {175},
    pages   = {103477},
    doi     = {10.1016/j.jdeveco.2025.103477}
}

@article{Lu2025RegressionMaps,
    author  = {Lu, Kerri and Kluger, Dan M. and Bates, Stephen and Wang, Sherrie},
    title   = {Regression Coefficient Estimation from Remote Sensing Maps},
    journal = {Remote Sensing of Environment},
    year    = {2025},
    volume  = {330},
    pages   = {114949},
    doi     = {10.1016/j.rse.2025.114949}
}

@article{rubin1976inference,
  title={Inference and missing data},
  author={Rubin, Donald B},
  journal={Biometrika},
  volume={63},
  number={3},
  pages={581--592},
  year={1976},
  publisher={Oxford University Press}
}

@article{carlson2025unifying,
  title={A unifying framework for robust and efficient inference with unstructured data},
  author={Carlson, Jacob and Dell, Melissa},
  journal={arXiv preprint arXiv:2505.00282},
  year={2025}
}

@article{rubin1974,
  title={Estimating causal effects of treatments in randomized and nonrandomized studies.},
  author={Rubin, Donald B},
  journal={Journal of Educational Psychology},
  volume={66},
  number={5},
  pages={688},
  year={1974},
  publisher={American Psychological Association}
}

@article{imbens2015causal,
  title={Causal inference for Statistics, Social, and Biomedical Sciences},
  author={Imbens, Guido W and Rubin, Donald B},
  journal={New York},
  volume={517},
  year={2015}
}

@misc{kennedy_semiparametric_2023,
	title = {Semiparametric doubly robust targeted double machine learning: a review},
	shorttitle = {Semiparametric doubly robust targeted double machine learning},
	url = {http://arxiv.org/abs/2203.06469},
	doi = {10.48550/arXiv.2203.06469},
	urldate = {2024-10-31},
	publisher = {arXiv},
	author = {Kennedy, Edward H.},
	month = jan,
	year = {2023},
	note = {arXiv:2203.06469},
}

@article{Hansen2013ForestChange,
    author  = {Hansen, Matthew C. and Potapov, Peter V. and Moore, Rebecca and Hancher, Matt and Turubanova, Svetlana A. and Tyukavina, Alexandra and Thau, David and Stehman, Stephen V. and Goetz, Scott J. and Loveland, Thomas R. and Kommareddy, Anil and Egorov, Alexey and Chini, Louise and Justice, Christopher O. and Townshend, John R. G.},
    title   = {High-Resolution Global Maps of 21st-Century Forest Cover Change},
    journal = {Science},
    year    = {2013},
    volume  = {342},
    number  = {6160},
    pages   = {850--853},
    doi     = {10.1126/science.1244693}
}

@article{Athey2026SurrogateIndex,
    author  = {Athey, Susan and Chetty, Raj and Imbens, Guido W. and Kang, Hyunseung},
    title   = {The Surrogate Index: Combining Short-Term Proxies to Estimate Long-Term Treatment Effects More Rapidly and Precisely},
    journal = {The Review of Economic Studies},
    year    = {2026},
    volume  = {93},
    number  = {4},
    pages   = {2284--2312},
    doi     = {10.1093/restud/rdaf087}
}

@article{Kallus2025Surrogates,
    author  = {Kallus, Nathan and Mao, Xiaojie},
    title   = {On the Role of Surrogates in the Efficient Estimation of Treatment Effects with Limited Outcome Data},
    journal = {Journal of the Royal Statistical Society Series B: Statistical Methodology},
    year    = {2025},
    volume  = {87},
    number  = {2},
    pages   = {480--509},
    doi     = {10.1093/jrsssb/qkae099}
}

@misc{DonahueEtAl2026, 
    title={Leveraging signals in expert trace data for interpretable complementarity},
    author={Kate Donahue and Alexander Idarraga and Rohan Alur and Gabriel Brat and Manish Raghavan},
    year={2026}
}

@article{PELLETIER2026103655,
    author  = {Johanne Pelletier and Mira Korb and Solomon Alemu and Manex B. Yonis and Travis J. Lybbert and Matthieu Stigler},
    title   = {Causal Inference with Predicted Outcomes: Correcting prediction error bias in satellite-based impact evaluation},
    journal = {Journal of Development Economics},
    year    = {2026},
    volume  = {179},
    pages   = {103655}
}

@misc{coqueret2026randomnesslargelanguagemodels,
      title={Randomness in large language models: What researchers need to know (and report)}, 
      author={Guillaume Coqueret and Joan Llull and Florian Oswald and Christophe Pérignon and Christoph Scheuch and Lars Vilhuber},
      year={2026},
      eprint={2607.24372},
      archivePrefix={arXiv},
      primaryClass={econ.GN},
      url={https://arxiv.org/abs/2607.24372}, 
}

@misc{BarrieEtAl24-LLMReplication,
  title={Replication for Language Models Problems, Principles, and Best Practice for Political Science},
  author={Christopher Barrie and Alexis Palmer and Arthur Spirling},
  year={2024},
  note={\nolinkurl{https://arthurspirling.org/documents/BarriePalmerSpirling_TrustMeBro.pdf}}
}

@article{Angrist2022,
author = {Angrist, Joshua D.},
title = {Empirical Strategies in Economics: Illuminating the Path From Cause to Effect},
journal = {Econometrica},
volume = {90},
number = {6},
pages = {2509-2539},
doi = {https://doi.org/10.3982/ECTA20640},
url = {https://onlinelibrary.wiley.com/doi/abs/10.3982/ECTA20640},
eprint = {https://onlinelibrary.wiley.com/doi/pdf/10.3982/ECTA20640},
year = {2022}
}

@article{Kruskal1977ThreeWayArrays,
    author  = {Kruskal, Joseph B.},
    title   = {Three-Way Arrays: Rank and Uniqueness of Trilinear Decompositions, with Application to Arithmetic Complexity and Statistics},
    journal = {Linear Algebra and its Applications},
    year    = {1977},
    volume  = {18},
    number  = {2},
    pages   = {95--138},
    doi     = {10.1016/0024-3795(77)90069-6}
}

@article{Allman2009Identifiability,
    author  = {Allman, Elizabeth S. and Matias, Catherine and Rhodes, John A.},
    title   = {Identifiability of Parameters in Latent Structure Models with Many Observed Variables},
    journal = {The Annals of Statistics},
    year    = {2009},
    volume  = {37},
    number  = {6A},
    pages   = {3099--3132},
    doi     = {10.1214/09-AOS689}
}
\endgroup

\clearpage
\appendix
\section{A Checklist for Observation with AI-Generated Variables}\label{app:checklist}

\begin{tcolorbox}[
    colback=forkbg, colframe=accent, boxrule=0.9pt, arc=2.5pt,
    left=8pt,right=8pt,top=6pt,bottom=6pt, before skip=4pt, after skip=10pt]
\textbf{Purpose.} This checklist helps authors report the measurement details a reader needs in order to assess AI-generated measures. We recommend either answering these questions explicitly in supplemental materials or using the checklist to confirm that its details are discussed somewhere in the article or replication materials. Answer to the best of your knowledge, and state \emph{``unknown''} where a detail is not available, which is common for variables obtained from an external source. If a question is irrelevant, provide justification. 
\end{tcolorbox}

\cksection{Provenance and Reproducibility}
\begin{checkitems}
  \item Did you construct the AI-generated variable(s), or did you obtain them from an external source? If external, what is the source?
  \item If you constructed the AI-generated variable(s), will the code, prompts, and model versions be released no later than the time of publication, along with the data where permissible?
  \item Do the replication files include the outputs returned by the model, so that the measurements can be reproduced without re-querying it?
\end{checkitems}

\cksection{Construct Definition}
\begin{checkitems}
  \item What is the definition of the construct(s) you measured with AI?
  \item Was any construct validation performed? If so, what evidence was used (\textit{e.g.,} variables the construct is shown to correlate with, or to remain distinct from)?
\end{checkitems}

\cksection{Validation Regime}
\cknote{Identify which of the three regimes you are in. Complete only the matching sub-section below.}
\begin{checkitems}
  \item Which validation regime applies: \textbf{(a) no validation data},
  \textbf{(b) representative validation data}, or \textbf{(c) validation data available, but not
  for the sample of interest}?
\end{checkitems}

\cksubsection{(a) No validation data}
\begin{checkitems}
  \item Why was obtaining validation data infeasible?
  \item What other evidence supports the validity of the AI-generated measure in the absence of a validation sample?
\end{checkitems}

\cksubsection{(b) Representative validation data}
\cknote{The validation labels are a random (conditional on observables) subsample of the
observations used to generate predictions.}
\begin{checkitems}
  \item How was the validation sample selected (\textit{e.g.,} simple random sample, or at random conditional on covariates)? If covariates were used, what are they?
  \item Is the probability that an observation is annotated known by design, or was it estimated (and if estimated, how)? Did every type of observation have a positive probability of being annotated?
\end{checkitems}

\cksubsection{(c) Validation data available, but not for the sample of interest}
\cknote{Ground truth exists, but it was collected in another sample (\textit{e.g.,} another region, period, or
population).}
\begin{checkitems}
  \item How do the two samples differ (\textit{e.g.,} in the distribution of predictions, covariates, geography, or time period)? Which forms of shift matter most in your application?
  \item Which stability assumption do you rely on---such as outcome stability or measurement stability---and what evidence supports it (\textit{e.g.,} comparisons of error rates, calibration, or predictive performance across the two samples)? If neither is defensible in your setting, say so.
\end{checkitems}

\cksection{Validation Dataset Description (if applicable)}
\begin{checkitems}
  \item How large is the validation sample, and when were the labels collected?
  \item What technology produced the validation labels (\textit{e.g.,} human annotators, instrument measurement)? Provide enough detail for a reader to understand how the measurement was implemented: for annotators, what training and instructions they were given; for instruments, the relevant technical specifications.
  \item Was a consistent rubric used to produce the validation labels?
  \item Was inter-annotator (or inter-instrument) agreement examined? If so, on how many observations, what was the agreement, and how were disagreements resolved?
\end{checkitems}

\cksection{Construction of the AI Predictions}
\begin{checkitems}
  \item What model(s) were used, and were they off-the-shelf or fine-tuned? Give model versions and citations, where the models can be obtained, and for API-served or hosted models the access date and version string.
  \item How were the model and the prompt selected? What alternatives were considered, and were the reported estimates examined for sensitivity to those choices?
  \item If a prompt was used, what was it? Provide the full text, along with the generation settings for a language model (\textit{e.g.,} temperature, top-$p$, maximum tokens, random seed, and number of samples drawn per observation).
  \item If training data were used to produce the predictions, how were the training data created? How large was the training sample? What were the relevant training details?
  \item How many observations have AI-generated measurements?
  \item Are the AI-generated measurements aggregated or otherwise transformed before use? If so, how, and at what level relative to the unit of analysis?
\end{checkitems}

\cksection{Use in Estimation}
\cknote{This checklist does not prescribe an estimator. The appropriate correction depends on the
estimand, on the role the measured variable plays, and on the validation regime above.}
\begin{checkitems}
  \item Does the AI-generated variable enter the analysis as an outcome, a regressor, an instrument, or in some other role?
  \item Do the reported estimates account for error in the AI-generated measurements? If so, what approach was used and what does it assume? If not, why is no correction needed for the estimand of interest?
\end{checkitems}

\end{document}